\documentclass[letterpaper]{article}
\usepackage[T1]{fontenc}

\usepackage{geometry}
\usepackage{setspace}

\usepackage[style = chem-acs, doi=true, articletitle=true]{biblatex}
\usepackage{url}
\usepackage{xcolor}
\usepackage{etoolbox}
\usepackage{graphicx}
\usepackage{enumitem}
\usepackage{subcaption}
\usepackage{listings}
\usepackage{amsmath}
\usepackage{lipsum}

\usepackage{authblk}
\author[1]{Natalie E. Hooven}
\author[1]{Arthur Y. Lin}
\author[1]{Charles H. Carroll}
\author[1,2,3]{Rose K. Cersonsky*}
\affil[1]{Department of Chemical and Biological Engineering, University of Wisconsin - Madison, Madison, WI, USA}
\affil[2]{Department of Materials Science and Engineering, University of Wisconsin - Madison, Madison, WI, USA}
\affil[3]{Data Science Institute, University of Wisconsin - Madison, Madison, WI, USA}

\title{Extrapolation of Machine-Learning Interatomic Potentials for Organic and Polymeric Systems}

\date{*Email: rose.cersonsky@wisc.edu}

\begin{document}

\maketitle

\begin{abstract}
Machine-Learning Interatomic Potentials (MLIPs) have surged in popularity due to their promise of expanding the spatiotemporal scales possible for simulating molecules with high fidelity. The accuracy of any MLIP is dependent on the data used for its training; thus, for large molecules like polymers and biomolecules, where ab initio training data is prohibitively difficult to obtain, it becomes necessary to use smaller, analogous chemical systems to construct MLIPs. Here, we perform a control study using $n = 1-8$ linear alkanes to determine when MLIPs trained on small molecules can accurately extrapolate to longer chains and more complex architectures. By combining MLIP performance analysis with environment‑resolved SOAP descriptors and Principal Covariates Classification, we show that reliable extrapolation emerges precisely when the relevant local chemical environments have converged between training and target systems. We further demonstrate that careful construction of the neighbor list substantially improves the learnability of intermolecular energetics, the component most critical for polymeric behavior. Together, these results provide a practical, data‑driven blueprint for designing transferable MLIPs for macromolecular materials, whether built from bespoke training data or adapted from emerging universal MLIP frameworks.

\end{abstract}


\section{Introduction}


In molecular simulation, the choice of potential energy model determines both the accuracy of predicted behavior and the computational time scales that can be achieved. Classical force fields have played a foundational role in polymer science for decades, enabling the study of thermodynamics, structure, and dynamics across a wide variety of molecular and polymeric systems. Their conceptual clarity, expressing molecular energetics through well‑defined bonded and nonbonded terms, combined with their computational efficiency, continues to make them indispensable for simulations that require long time scales, large system sizes, or broad chemical screening.

At the same time, advances in electronic‑structure theory, machine learning, and high‑performance computing have created new opportunities at the interface of accuracy and efficiency. Machine‑learned interatomic potentials (MLIPs) offer one such opportunity: by learning from quantum‑chemical reference data, they can capture complex many‑body interactions without requiring an explicit decomposition into fixed functional forms\cite{bartok_representing_2013, batatia_mace_2022, behler_generalized_2007, cheng_cartesian_2024, batzner_e3-equivariant_2022, lindsey_chimes_2017, lin_expanding_2024, lin_anisoap_2025}. MLIPs have recently demonstrated strong performance across problems ranging from ionic transport\cite{winter_simulations_2023, goodwin_transferability_2024}, condensed-phase phenomena\cite{cheng_ab_2019, zhang_phase_2021}, and electronic properties\cite{grisafi_transferable_2019}, motivating interest in when and how these models can complement existing force‑field approaches--particularly for systems where non-additive electronic effects or chemical heterogeneity play an important role.

However, using MLIPs for large molecules such as polymers or biomolecules remains challenging. High‑quality quantum chemical data are difficult to obtain for such macromolecules\cite{xu_new_2022}, and transferability across chain length or architecture is not guaranteed or well-understood. This has led to increasing interest in strategies where smaller, chemically related molecules act as surrogates for constructing potentials that are then applied to larger targets\cite{jacobson_transferable_2022, mohanty_development_2023, unke_biomolecular_2024}. Although intuitive, the limits of such extrapolation have not been systematically quantified.

Universal or ``foundational'' machine-learned interatomic potentials (UMLIPs) have recently emerged as a promising direction, offering broad chemical coverage from large, diverse training sets\cite{mazitov_massive_2025, batatia_foundation_2024, radova_fine-tuning_2025, yuan_foundation_2025}. These models reduce the need for system‑specific quantum data and have shown strong performance across many small‑molecule and materials benchmarks. However, due to their recency, their behavior for macromolecular systems remains largely untested. Moreover, even when UMLIPs provide a reliable baseline, fine‑tuning or system‑specific adaptation is often needed for application-specific studies. And thus, understanding which molecular environments must be represented is still quintessential, regardless of whether one uses a bespoke MLIP or a UMLIP.

In this work, we provide a control study of this question. Using linear alkanes with $n = 1-8$ as a model system, we examine when MLIPs trained exclusively on small molecules can accurately extrapolate to longer chains and more complex architectures. By combining MLIP performance analysis with environment‑level descriptors and classification tools, we identify the molecular environments that govern extrapolation accuracy and show how neighbor‑list design influences the learnability of inter‑ versus intramolecular interactions. These results form a practical blueprint for building transferable MLIPs for macromolecular systems, in a way that complements, and in some cases extends, the capabilities of traditional force‑field approaches.

\section{Methods}\label{sec:methods}

\subsection{Data generation}

\subsubsection{Training sets}
Each MLIP was trained on a single $n=1-8$ $n$-polyalkane dataset. To limit confounding factors in our analyses, datasets were constructed at similar phase, pressure, and temperature conditions. Each dataset consisted of $N=64$ molecules, simulated at $T=300$K with the volume corresponding to NIST fluid properties' density for each alkane at 5 MPa and 300K\cite{linstrom_nist_1997}. At these temperature and pressure conditions, each alkane is in a liquid state, aside from methane, which is a supercritical liquid. 
Initial configurations of $n=3-8$ polyalkanes were obtained from the Cambridge Structural Database\cite{groom_cambridge_2016}. For methane and ethane, initial configurations were generated using RDKit\cite{greg_landrum_rdkitrdkit_2025} on a cubic lattice with the appropriate volume. Each initial configuration was relaxed to an unordered, liquid state  using isobaric-isothermal (NPT) molecular dynamics simulations at low pressure ($P=0.5$ MPa, $T=300$K). These NPT simulations were performed with a Berendsen barostat and thermostat, a timestep of 0.2 femtoseconds as well as periodic boundary conditions, and were run until the reference experimental density was obtained for each polyalkane. 

For $n=7,8$ datasets, further relaxation at large box volume (with $P=0.01$ MPa, $T=300$K) and geometry optimization was necessary in order to remove initialization artifacts. After these systems were randomized, we repressurized these two datasets to 300K and 5 MPa with the OPLS-AA forcefield until they reached the appropriate volume \cite{jorgensen_development_1996}using periodic boundary conditions, a timestep of 1 femtosecond, and a Nose-Hoover thermostat and barostat. Initial velocities were generated using a Maxwell-Boltzmann sampling at 300 K without fixing a random seed. With n = 8, some initialization artifacts still remained, so we computed structure optimization in ASE using a BFGS algorithm and fmax=0.05 with a DFTB+ calculator using a Conjugated Gradient driver to anneal the system. 

After the appropriate box volumes were reached, NVT simulations were then conducted at 300K. These production simulations were conducted using DFTB+, initially using a grid of (1, 1, 1) k-points with a Berendsen thermostat, an initial stochastic Maxwell-Boltzmann velocity distribution, a timestep of 0.2 femtoseconds, as well as periodic boundary conditions, and were run until potential energies had converged. For our purposes, an equilibrium cutoff was defined as where the running average of the total energy did not fluctuate by more than 0.1 eV for 2,000 consecutive frames in the NVT simulation. We then selected 1000 frames from the subsequent simulation at random to form testing sets, and used farthest-point sampling\cite{goscinski_scikit-matter_2025, imbalzano_automatic_2018, cersonsky_improving_2021} on Smooth Overlap of Atomic Positions (SOAP\cite{bartok_representing_2013}) vectors to select a mutually-exclusive 10,000 frames to constitute our training sets. 

\subsubsection{Additional testing sets}
Decane, dodecane, cyclohexane, 4-propylheptane, and 3,3-diethylpentane, configurations were sampled using the OPLS all-atom (AA) forcefield\cite{jorgensen_development_1996}. Decane and dodecane were simulated with 48 molecules, and the others with 64 molecules, in order to keep computational overhead consistent. The molecules were initialized on a lattice with 10 $\text{\AA}$ spacing and with each dimension of each lattice position randomly perturbed by -5 to 5 $\text{\AA}$ sampled from a uniform distribution. We equilibrated the systems with respect to the average radius of gyration, which quantifies the average size of the alkane conformations. The radius of gyration of an alkane, $R_g$, is given by
$R_g=\sqrt{\frac{\sum_im_ir_i^2}{\sum_im_i}},$
where the sums are over atoms $i$ in the alkane, $m_i$ is the atomic mass, and $r_i$ is the displacement relative to the alkane center of mass. We used pyMBAR \cite{chodera_simple_2016} to calculate the correlation time of the radius of gyration time series. We equilibrated each system for 100 correlation times. Figure \ref{si:rg} demonstrates the equilibrium radius of gyration of the alkanes. Once equilibrated, the systems were run for 1 ns. All classical simulations were run in LAMMPS \cite{thompson_lammps_2022} using forcefield parameters generated by ligpargen\cite{jorgensen_potential_2005, dodda_114cm1a-lbcc_2017, dodda_ligpargen_2017} in an NPT ensemble at 300 K and 5 MPa using a Nose-Hoover thermostat and barostat with loose couplings of 1 $\text{ps}^{-1}$ each and a timestep of 1 fs.

\subsubsection{Energy and force evaluations}
For all frames in the training and testing data, we recomputed the energies and forces with DFTB+ using self-consistent charges and the mio Slater-Koster files\cite{elstner_self-consistent-charge_1998}  using a Monkhurst-Pack grid-point interpolation (3, 3, 3). We utilized a high-throughput workflow with \texttt{signac}\cite{adorf_simple_2018} and \texttt{signac-flow}\cite{ramasubramani_signac_2018}.

To construct values for inter- and intra-molecular energies and forces, we isolated the individual molecules into separate ASE-atoms objects and recomputed energies and forces without consideration of periodic boundary conditions. This yielded
$$E_{intermolecular, \ bulk}=E_{total, \ bulk}-E_{intramolecular}$$ 
where $E_{intramolecular}\equiv\sum_{i=1}^{64} E_{i}$, with $E_{i}$ the total energy of the $i^{th}$ single molecule.

\subsection{Machine-learned potentials and analyses}

\subsubsection{MACE MLIPs}

For every MLIP, we solely used the aforementioned training sets during construction, and trained each model as one normally would for a non-extrapolative task. 
When training solely on intramolecular quantities (Fig.~\ref{si:results-intra}), we down-selected from the 640,000 single-molecule frames to 10,000, choosing one random molecule from each training frame. Testing sets were kept consistent, with 64,000 single molecule frames in total.

The MACE\cite{batatia_mace_2022} model (Higher‑Order Equivariant Message Passing Neural Network) combines equivariant graph neural network design with explicit many‑body message passing. Unlike most standard MPNNs that rely on only two‑body (pairwise) interactions and require many message‑passing layers to build complexity, MACE constructs higher‑body (up to four‑body) equivariant messages in a hierarchical fashion. This allows the network to achieve high representational power with just two message‑passing iterations, substantially reducing depth, improving parallelism, and maintaining scalability while reaching or surpassing state‑of‑the‑art accuracy on benchmarks such as rMD17, 3BPA, and AcAc\cite{batatia_mace_2022}. By building hierarchical tensor products of lower‑order features and symmetrizing via generalized Clebsch–Gordan coupling, MACE efficiently implements these many‑body messages without exponential cost.  This induces equivariant feature tensors, which are linearly combined (with weights dependent on element type and irreducible representation indices to produce message updates. 
For our MACE-based potentials, we use the software \texttt{v.0.3.13} and a standard set of model hyperparameters, as shown in the example script in the Appendix.

\begin{figure*}[ht!]
    \begin{subfigure}[t]{0.45\linewidth}
    \includegraphics[width=0.9\linewidth]{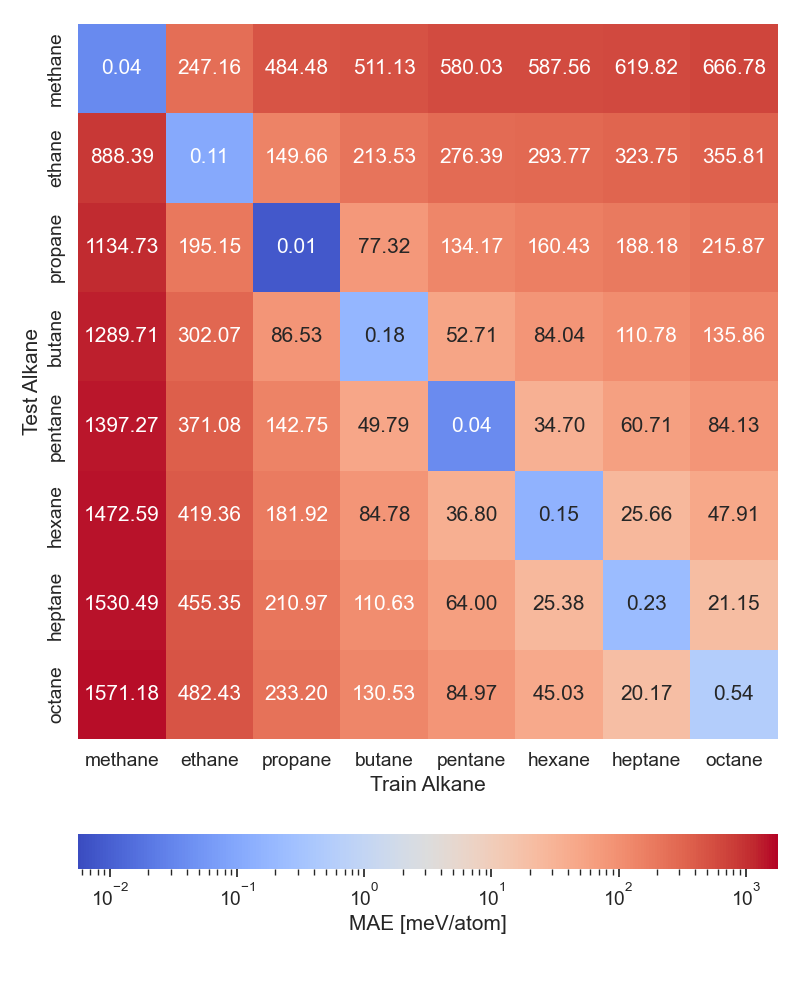}
    \caption{Results of predicting per-configuration energies (normalized by the number of atoms) using MACE MLIPs for different train (x-axis) and test (y-axis) pairings, respectively. Color indicates the mean-absolute-error (MAE) in units of meV/atom, respectively.}
    \label{fig:energies}
    \end{subfigure}\hfill\begin{subfigure}[t]{0.45\linewidth}
    \includegraphics[width=0.9\linewidth]{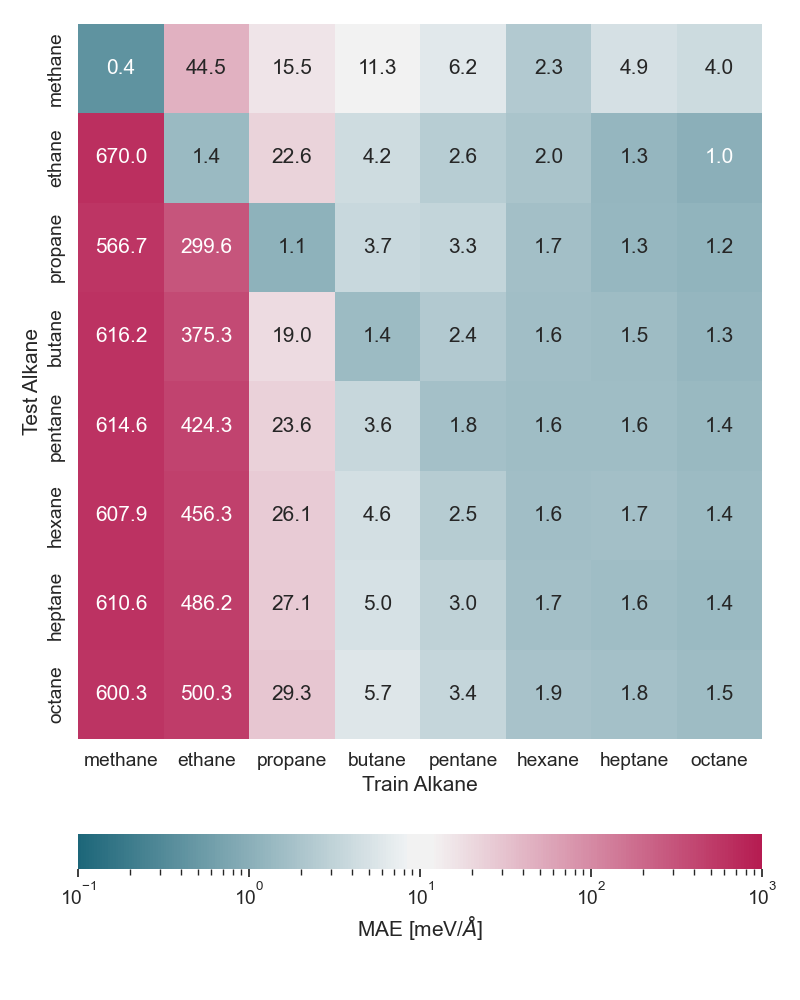}
    \caption{Results of predicting atomic force-vectors using the MACE MLIPs for different train (x-axis) and test (y-axis) pairings, respectively. Errors correspond to the average magnitude of vectorial error across the testing set. Color indicates the mean-absolute-error (MAE) in units of meV/\AA, respectively.}
    \label{fig:forces}
    \end{subfigure}
    
    \begin{subfigure}[b]{0.45\linewidth}
    \includegraphics[width=0.9\linewidth]{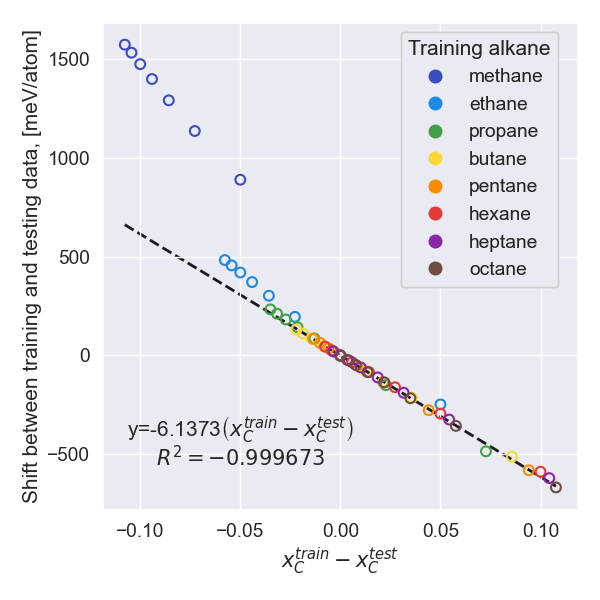}
    \caption{The relationship between the shift in composition $(x_C^{\text{train}} - x_C^{\text{test}})$ and the shift in energies for extrapolated MLIPs. The models trained on the methane, ethane, and propane datasets show a divergence from this proportionality, as they are not necessarily considered ``well-conditioned.''}
    \label{fig:mean-shift}
    \end{subfigure}\hfill\begin{subfigure}[b]{0.45\linewidth}
    \includegraphics[width=0.9\linewidth]{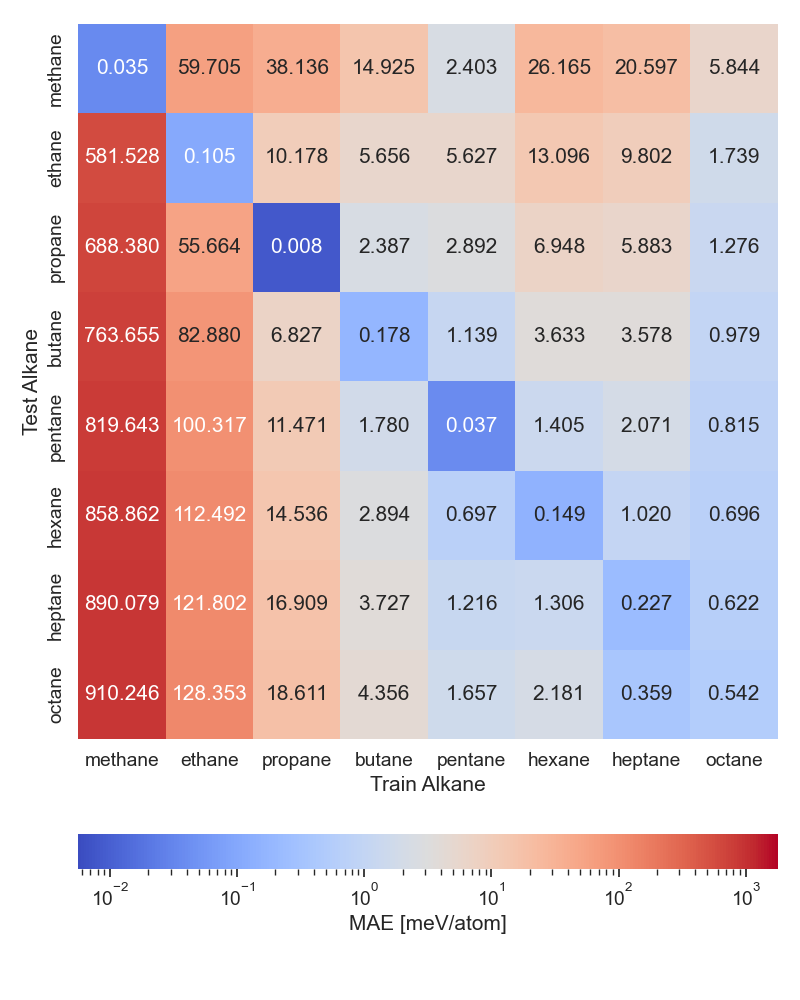}
    \caption{Shifted energy errors for MACE energies, accounting for the learnable shift in Fig.~\ref{fig:mean-shift}.}
    \label{fig:mean-shift-correction}
    \end{subfigure}
    \caption{Results of MACE potential building on the total potential energy on $n=1-8$ alkanes.}
    \label{fig:results}
\end{figure*}

\subsubsection{SOAP Analysis}
Smooth Overlap of Atomic Positions (SOAP) vectors were computed using the \texttt{featomic}\cite{bigi_metatensor_2025} and \texttt{librascal} \cite{musil_efficient_2021} packages with an interaction cutoff of $10$\AA~(for Fig.~\ref{fig:map}, Fig.~\ref{si:pcovc_pcova}) and $7$\AA~(for Fig.~\ref{fig:soap_ridge}, Fig.~\ref{si:rel_soap}). In all cases, we used the GTO basis set with 8 radial components and 4 angular components. Principal Covariates Classification (PCovC, \cite{jorgensen_interpretable_2026}) was computed using v0.3.2 of \texttt{scikit-matter}\cite{goscinski_scikit-matter_2025}, using as $X$ the aforementioned SOAP vectors of the training set, a mixing parameter of $0.5$, a logistic regression classifier, and oligomer length $n$ as our class value. Each PCovC model was computed solely for like-environments, such as hydrogen, CH$_2$, or tertiary carbon environments (Fig.~\ref{fig:map}, Fig.~\ref{si:pcovc_pcova}). Prior to fitting and analysis, feature vectors and energies were appropriately scaled and centered using \texttt{scikit-matter}\cite{goscinski_scikit-matter_2025} \texttt{StandardFlexibleScaler}. For regression (Fig.~\ref{fig:soap_ridge}), we used \texttt{scikit-learn}\cite{pedregosa_scikit-learn_2011} \texttt{RidgeCV} with \texttt{cv=5} and a negative mean squared error scorer. 20 regularization constants ($\alpha$) were considered for RidgeCV using a logarithmic grid.

\section{Results}\label{sec:results}
\subsection{Extrapolating energies and forces}

Fig.~\ref{fig:results} shows the results of using trained MLIPs to predict the energies and forces on different oligomeric chain lengths. On first glance, the energy-fitting results of Fig.~\ref{fig:energies} tell a discouraging story -- each training set is unable to predict the energies of the other polyalkanes, with a MAE well-above chemical accuracy (which we define as 1meV/atom\cite{zuo_performance_2020}). However, the force results (Fig.~\ref{fig:forces}) imply something much different, in line with our chemical intuition -- at larger chain lengths, we can reasonably extrapolate the energetics of one system to the next (and even those below). 

So, why do these two results diverge? A first intuition would be data centering -- machine-learning potentials are typically trained to predict the \emph{fluctuations} in molecular energetics, not their absolute value. In other words, the majority of MLIP methods and software first subtracts the mean of the training targets before determining weights and biases. This means that the predicted energy $E_{pred}$ is ostensibly $$E_{pred} = E_{MLIP} + b,$$ where $b$ is a ``baselining constant,'' or the energetic values shared by all training points. This is verified by the strong colinearity between our predicted and true values, as shown in Fig.~\ref{si:colinear}.
A simple explanation of this ``mean-shift'' would be the shift in atomic energies, $$E_ {atom} = n_C E_C + n_H E_H = N (x_C E_C + x_H E_H),$$ which, when learning per-atom energies $E / N$, would yield  $$b \equiv (x_C^\text{train} - x_C^\text{test})(E_C - E_H).$$ However, this is not the case here, where a shift due solely to atomic contributions would lead to a much larger value than the observed shift. This presents an issue for extrapolating energetics from one system to another, where the only information known \textit{a priori} may be the difference in underlying chemistry, and where ``transfer learning'' may be infeasible due to the lack of underlying training data. Thus, we first determine if $b$ is a linear function of our composition ($x_C$, $x_H$), as this would still allow for the extrapolation of one system to another without prior computation of energetics or configurations.

For models trained on $n=4-8$ alkanes, there exists a strong linear proportionality (Pearson correlation coefficient = 0.9997, Fig.~\ref{fig:mean-shift}) between the compositional shift and the mean-shift of the datasets. This suggests that these mean-shifts are learnable parameters, even if not obviously grounded in physical values such as $E_{atom}$. To test this, we determined a slope to the line in Fig.~\ref{fig:mean-shift} via non-regularized linear regression, omitting from training the values obtained from models built on methane (blue), ethane (light blue), and propane (green), which are considered unsuccessful models. Accounting for the large co-linearity of the mean-shift and compositional change, many of the error predictions are improved, yet there exist some residual errors (Fig.~\ref{fig:mean-shift-correction}). This suggests that the change in conformational distribution contributes non-negligibly to the mismatch in energies.

However, the results of Fig.~\ref{fig:forces} still demonstrate that there exist two critical lengths of polyalkane for extrapolation: 1) butane, where we see a general reduction in force error magnitude from 20-30meV/\AA~to 3-6meV/\AA~(omitting methane), and 2) hexane, after which we see little improvement in force error with additional molecular units. The change in errors from propane to butane follows our chemical intuition -- prior to this chain length, there is no sampling of the dihedral rotation possible in these molecules, and thus any system ($n=4-8$) that contains these interactions would be ill-represented by such models.

But what makes hexane special? To understand this, we can look at the higher-dimensional representation of the interaction environments shared and distinct within the training sets. While we could arguably use the learned features of the MACE MLIPs, we instead employ a preceding, related technology,  the Smooth Overlap of Atomic Positions (SOAP\cite{bartok_representing_2013}), which are deterministic in their description of atomic environments without the implications of training on one dataset versus another . Both MACE and SOAP vectors compute the many-body expansion of atomic interactions, and thus we can infer some behaviors of MACE based on a SOAP representation. We construct our SOAP descriptors based on the distance of the MACE perceptive field (10.0\AA). 

After we compute the SOAP vectors for each training set, a standard analytical task is to look at the Principal Components Analysis (PCA) of all vectors, which would demonstrate the diversity of environments. However, here we are particularly interested in how the environments within these datasets differ, which is not the underlying mathematical motivation of PCA, and this aspect can be largely lost in PCA (and other unsupervised) maps. So, we instead employ hybrid supervised-unsupervised analyses called Principal Covariates Classification (PCovC)\cite{jorgensen_interpretable_2026, helfrecht_structure-property_2020, goscinski_scikit-matter_2023, goscinski_scikit-matter_2025}, which computes the two-dimensional projection that both retains PCA-like variance and best separates the data based on assigned labels (here the number of carbons in the dataset backbone). Here we focus on the three-body interactions of the CH$_2$ terms, which demonstrate the greatest statistical difference from dataset to dataset, with PCA and PCovC maps of remaining environments in Fig.~\ref{si:pcovc_pcova}. 

\begin{figure}[ht!]
\centering 

    \includegraphics[width=0.75\linewidth]{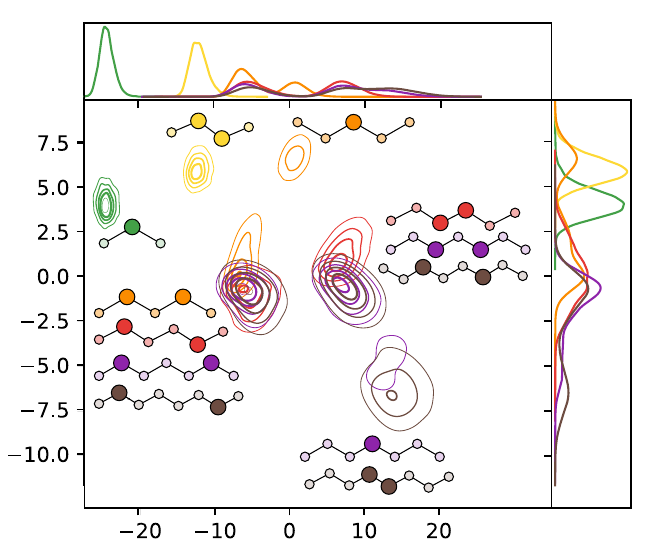}
    \caption{Distribution of CH$_2$ environments for the different training sets, mapped via Principal Covariates Classification (PCovC\cite{jorgensen_interpretable_2026}). The axes are directions in latent space that demonstrate the highest degree of diversity in the CH$_2$ environments and separability between the datasets. Points that are overlapping are considered ``indistinguishable,'' in that an ML model could not tell whether the environment sits in one alkane length or another. The map demonstrates the convergence of environments with increasing $n$. Different alkanes are denoted in different colors (green - propane, yellow - butane, orange - pentane, red - hexane, purple - heptane, and brown - octane), with the corresponding CH$_2$ environment enlarged for each cluster in the map.}
    \label{fig:map}
\end{figure}

From this analysis, we see that the convergence in MLIP error corresponds with the sampling of the CH$_2$ environments three carbons from the end, and the addition of further internal environments within heptane and octane leads to diminishing returns in terms of force approximation. This demonstrates that determining the minimal chain length required for MLIP construction (or molecular fragmentation\cite{gordon_fragmentation_2012,jacobson_transferable_2022, mohanty_development_2023,unke_biomolecular_2024}) amounts to determining the convergence of the majority of molecular environments, a task that can be accomplished via classification models or similar unsupervised tasks on single-molecule data.

\subsection{Separating the different contributions to total energies}
However, predicting the total energies and forces may not be sufficient for a usable potential model, as the hierarchical nature of interactions within these systems decides much of their fundamental behavior\cite{dral_hierarchical_2020}. Conceptually, and in classical force fields,, we consider potential energies emerging from three distinct classes of interactions: $$E = E_{atom} + E_{intramolecular} + E_{intermolecular},$$ as these terms have significantly different magnitudes, difficulties in calculating, and difficulties in terms of ``learnability.'' $E_{atom}$ is the largest component but arguably trivial to compute. We can group $E_{atom}$ with intramolecular interactions to obtain an $\tilde{E}_{intramolecular}=E_{atom}+E_{intramolecular}$ 
by sampling single molecule configurations; this is likewise a large component of $E$. As shown in Fig.~\ref{si:results-intra}, extrapolating $\tilde{E}_{intramolecular}$ follows the same lessons learned in predicting $E$ -- both in terms of the necessary molecule lengths for extrapolation and the ML baselining leading to divergent narratives for energy- and force-fitting.

The element of this energetic expansion that requires the greatest nuance is the \emph{intermolecular} interactions $E_{intermolecular}$, which are comparatively weak, longer-range, and computationally expensive to calculate using methods from quantum chemistry\cite{parker_levels_2014}. And, when training on total energies, the resultant errors often overshadow these small contributions\cite{dral_hierarchical_2020}, losing interactions that are decisive in many critical scientific problems. 

MACE itself (v.0.3.13) is not natively suited for directly learning these interactions, nor are atom-centered expansions that implicitly or explicitly focus on the nearby atoms in constructing their MLIP\cite{cersonsky_data-driven_2023}. To demonstrate this, we again turn to the SOAP\cite{bartok_representing_2013} framework, and show how careful attention to weighting interactions can yield better fidelity for intermolecular models. 
Following the same procedures in terms of training/testing splits, we construct cross-validated Ridge Regressions on SOAP representations for the same $n=1-8$ $n$-polyalkane datasets used for MACE MLIP construction. SOAP vectors were calculated with a cutoff of 7\AA~for atomic environments and averaged for each frame such that for the 10,000 training frames and 1,000 testing frames, we have 10,000 and 1,000 corresponding SOAP vectors; we refer to these initial SOAP calculations as the ``total'' SOAP vectors, or $X^\text{total}$. 

\begin{figure}[ht!]
    \begin{subfigure}[t]{0.48\linewidth}
    \includegraphics[width = \linewidth]{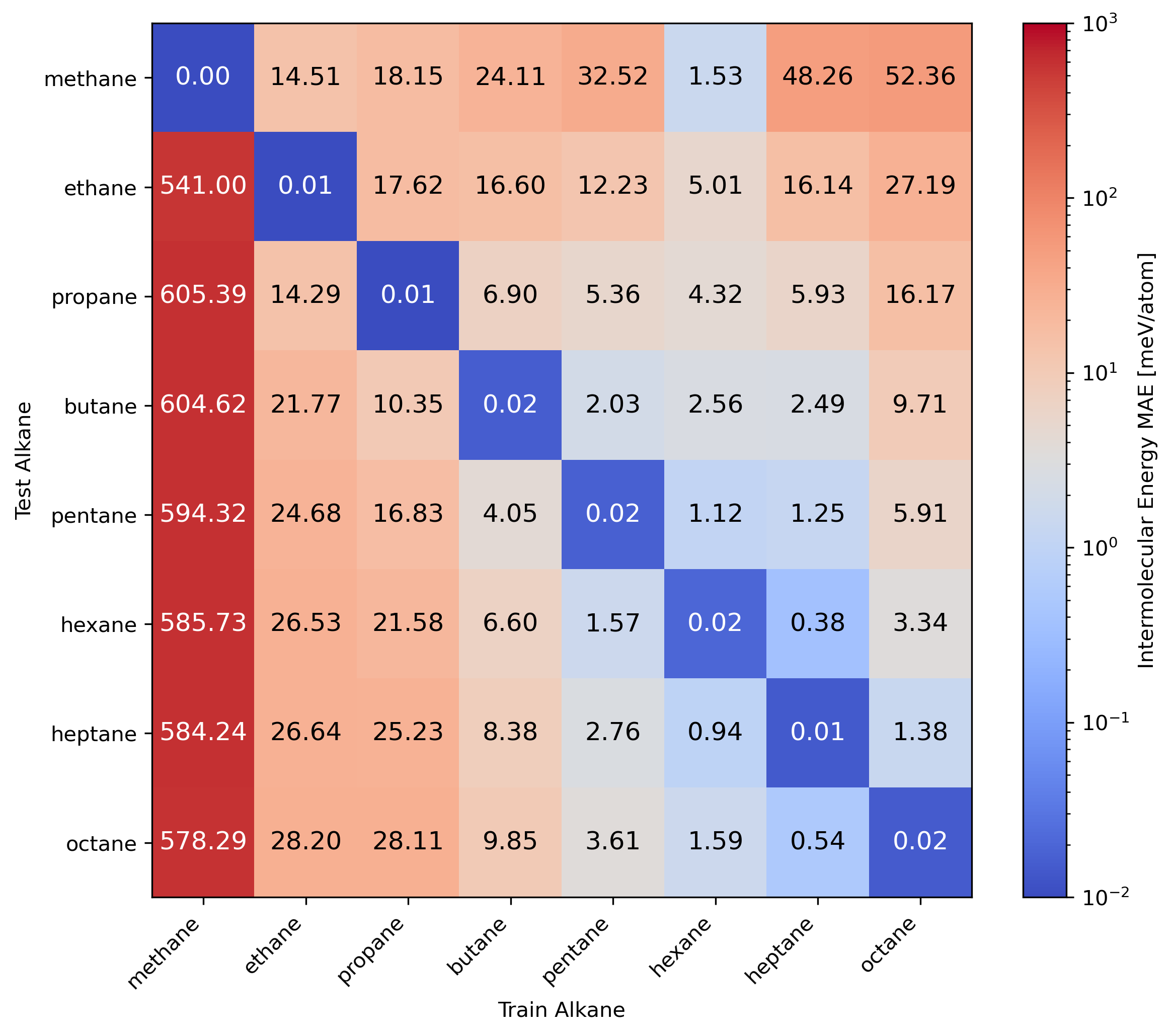}
    \caption{Regressions using total SOAP vectors $X^\text{total}$.}
    \label{fig:soap_baseline}
    \end{subfigure}
    \begin{subfigure}[t]{0.48\linewidth}
    \includegraphics[width = \linewidth]{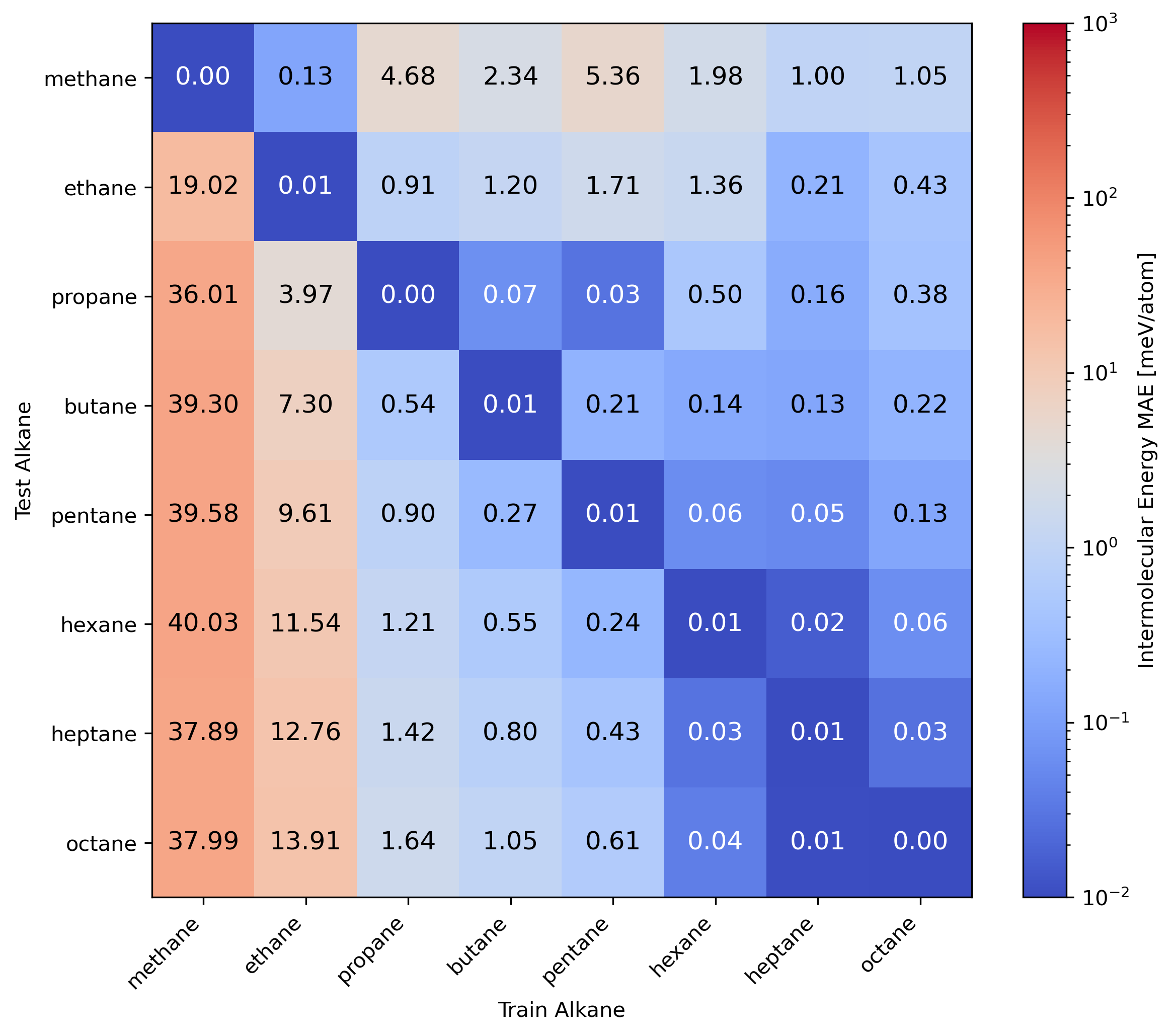}
    \caption{Regressions using the far-sighted SOAP vectors $X^{fs}$.}
    \label{fig:soap_partitioned}
    \end{subfigure}
    \caption{Results of SOAP-Ridge MLIPs trained on intermolecular potential energy using different SOAP representations.}
    \label{fig:soap_ridge}
\end{figure}

Similar to the initial results from MACE in Fig. \ref{fig:energies}, the results of using a $E_\text{intermolecular}\approx X^\text{total}w$, shown in Fig. \ref{fig:soap_baseline}, are discouraging. Extrapolation, with a field-standard threshold MAE of 1 meV/atom \cite{zuo_performance_2020}, is only achieved three times (by hexane on heptane, and heptane on hexane/octane), even despite the considerably smaller magnitude of the intermolecular energetics, as compared to the total energies regressed earlier. However, we can adopt lessons from learning the lattice energy of molecular crystals\cite{cersonsky_data-driven_2023} -- SOAP representations, and representations that are implicitly or explicitly ``near-sighted''\cite{batatia_mace_2022, drautz_atomic_2019, han_deep_2018} will, by design, consider the contributions of nearby atoms more strongly than those further away. This will overshadow the intermolecular environments which contribute to $E_{intermolecular}$, limiting our accuracy in extrapolation.

\begin{figure*}
    \centering
    \begin{subfigure}{0.75\linewidth}
    \includegraphics[width=\linewidth]{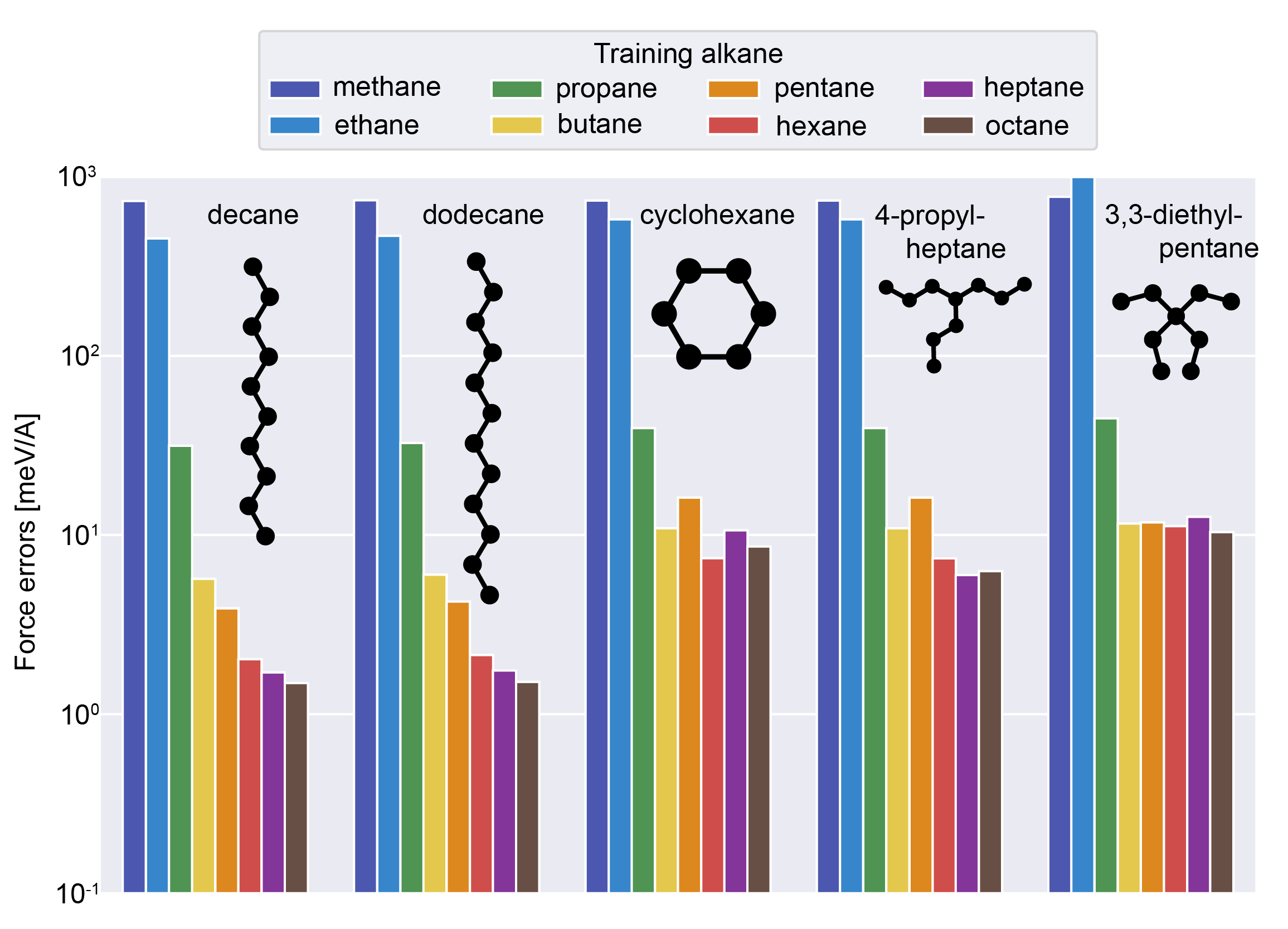}
    \caption{Force errors resulting from testing n=1-8 trained MLIPs on longer chain lengths and architectures. Color of the bar plot denotes the training set used, and each set of bars refers to a different test molecule.}
    \label{fig:specials}
    \end{subfigure}
    \begin{subfigure}[t]{\linewidth}
        \includegraphics[width=0.3\linewidth]{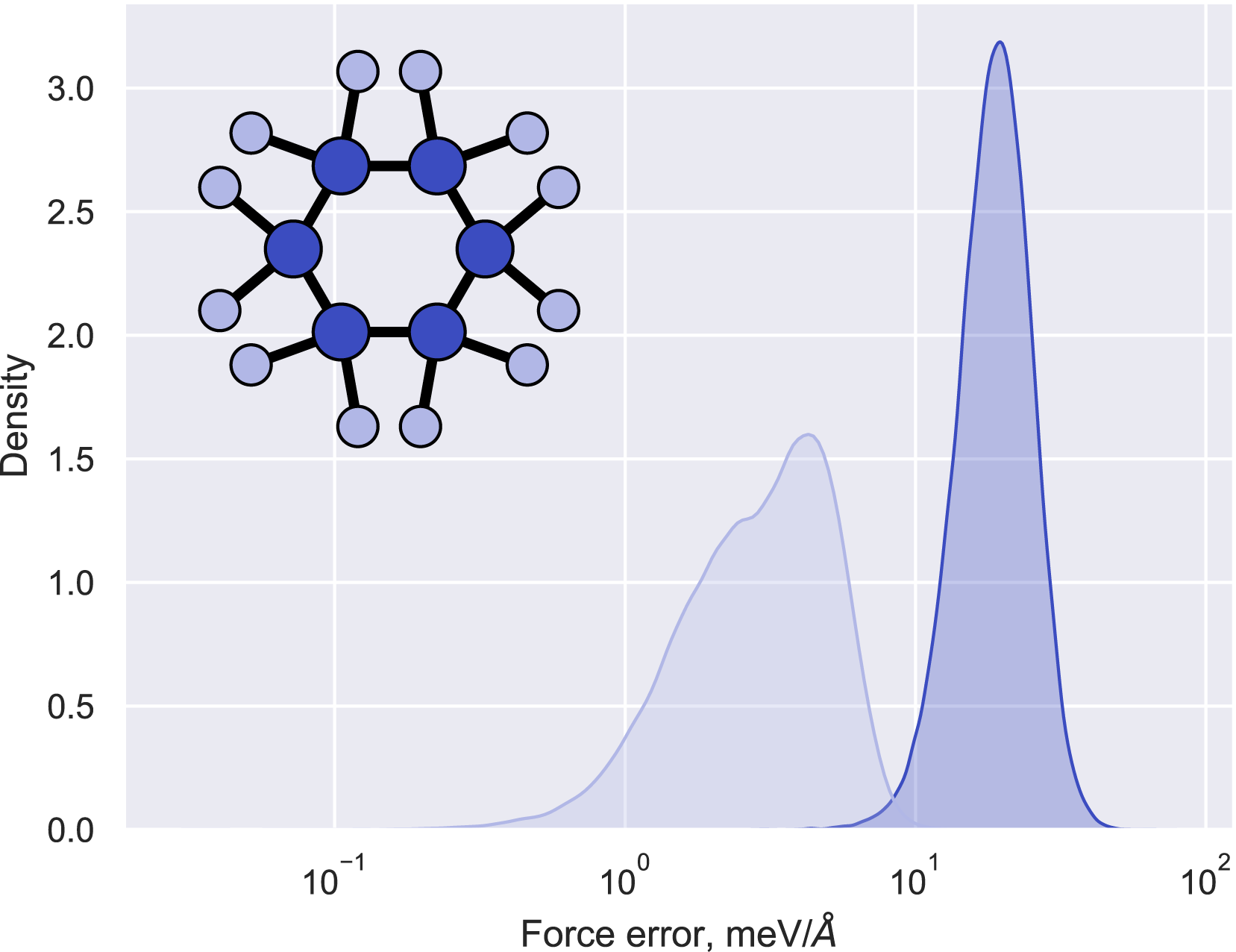}
        \includegraphics[width=0.3\linewidth]{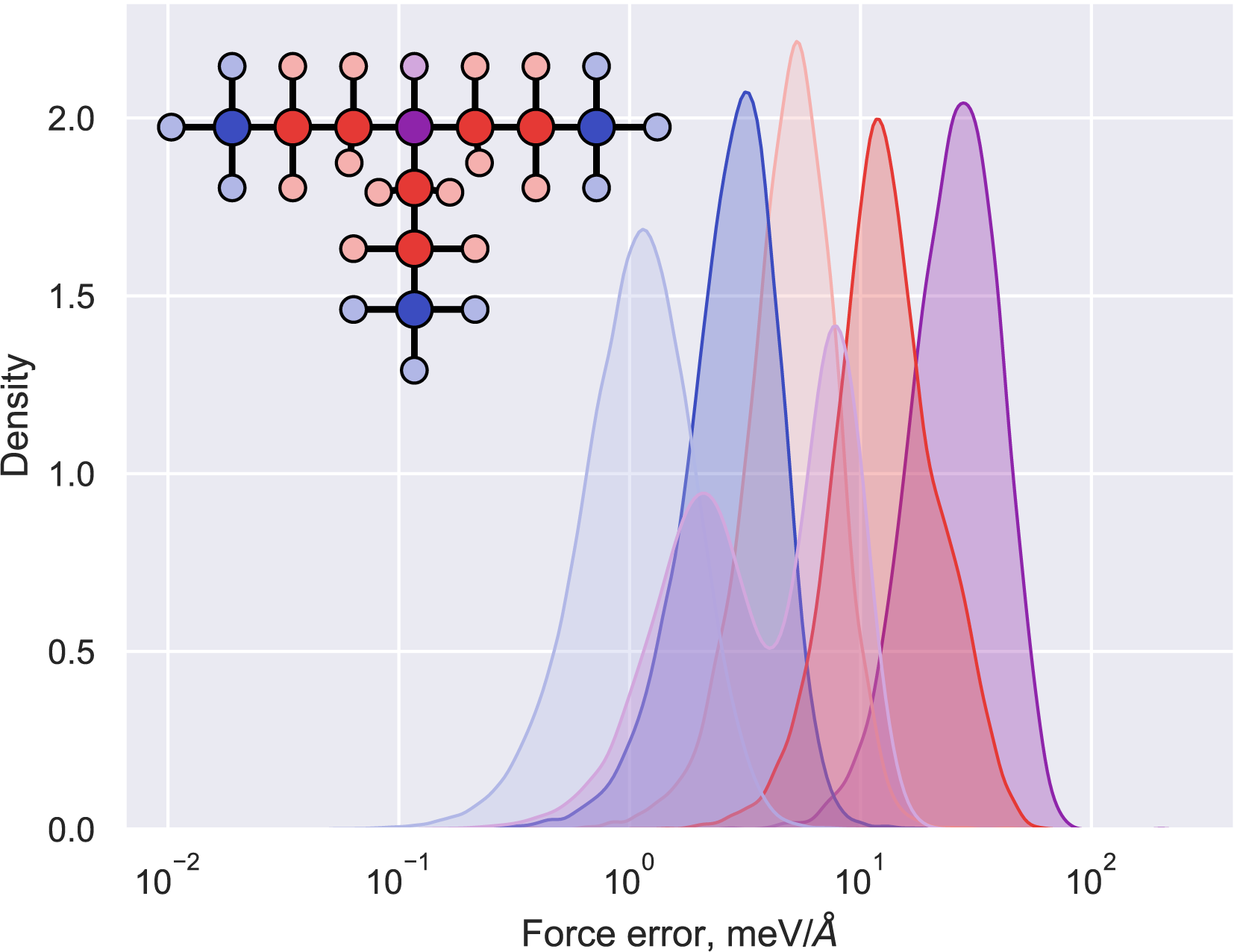}
        \includegraphics[width=0.3\linewidth]{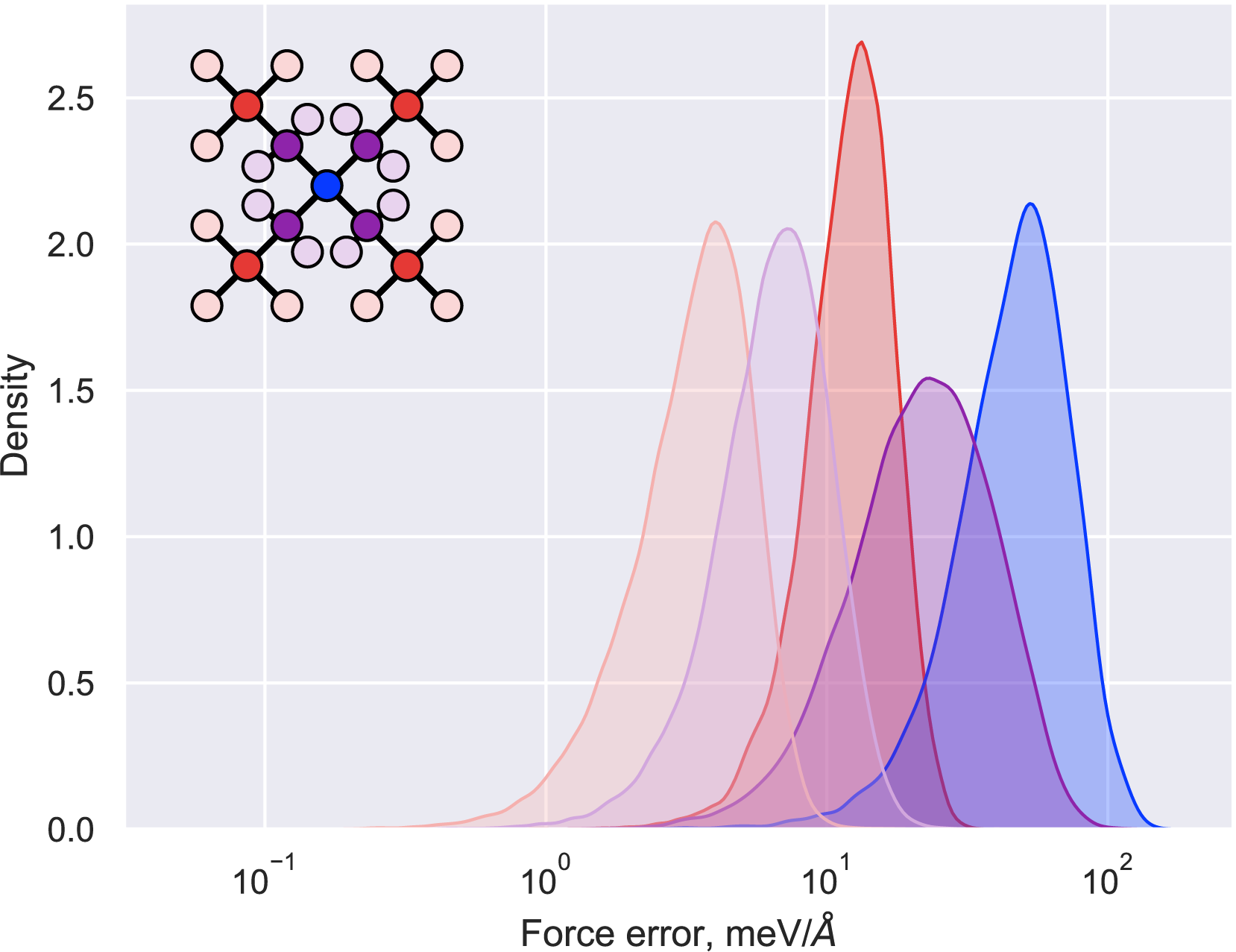}
        \caption{Distribution of force errors for cyclohexane (left), 4-propylheptane(center), and 3,3-diethylpentane (right), as predicted by a model trained on octane. Colors indicate the atoms within the molecule. For every carbon, the corresponding hydrogens are in the same hue, with a lighter tint.}
        \label{fig:kde}
    \end{subfigure}
    \begin{subfigure}{0.75\linewidth}
    \includegraphics[width=\linewidth]{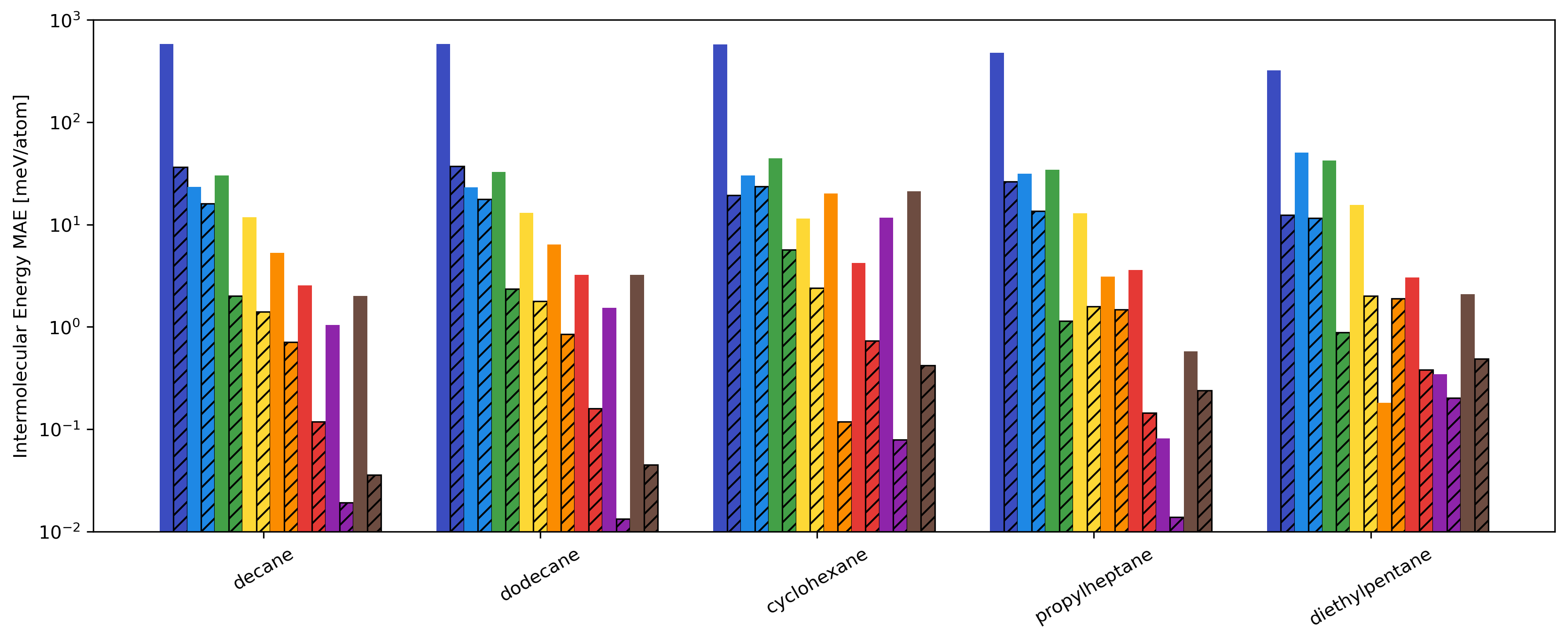}
    \caption{Intermolecular energy errors resulting from testing n=1-8 trained SOAP-Ridge MLIPs on longer chain lengths and architectures. Color of the bar plot denotes the training set used, and each set of bars refers to a different test molecule. Hatched bars indicate MLIPs constructed using far-sighted SOAP vectors $X^{fs}$, solid bars indicate MLIPs using total SOAP vectors.}
    \label{fig:specials_energies}
    \end{subfigure}
    \caption{Force and energy errors for decane, dodecane, and the nonlinear alkanes.}
\end{figure*}

We can thus construct a ``far-sighted'' SOAP vector $$X^{fs} = X^{total} -\frac{1}{64}\sum_{i=1}^{64}X_{i}$$ where $X_i$ is the averaged SOAP vector of the atomic environments in the $i$\textsuperscript{th} molecule within the bulk frame. Effectively, this representation re-weights the feature space by reducing (or eliminating) the influence of contributions from intramolecular environments on the SOAP vector. As shown in Fig. \ref{fig:soap_partitioned}, using this far-sighted representation results in a significant improvement in the number of extrapolative regimes achieved by SOAP-Ridge MLIPs in comparison to the results in Fig. \ref{fig:soap_baseline}. Similar trends are observed as in Fig. \ref{fig:forces}: steep improvements in MAEs are observed between ethane and propane/butane, and once again from pentane to hexane. This follows the convergence in chemical environments (Fig. \ref{fig:map}, Fig. \ref{si:pcovc_pcova}). Furthermore, these results demonstrate that careful consideration of MLIP features can improve performance for targets other than the total potential energy of a system.

\subsection{Beyond n=1-8 linear alkanes}
It is worth inspecting the limitations of these models when considering longer molecules or more complex hydrocarbon architectures. As shown in Fig.~\ref{fig:specials}, decane (n=10) and dodecane (n=12), the trends follow those similar to the shorter oligomers within the series: we see a reduction in error when moving from propane to butane (30meV/\AA~to 5-6meV/\AA), and a saturation near 1.5-2 meV/\AA~ once training on hexane through octane. This supports our earlier observations considering the saturation of interaction environments, and suggests that these trends should continue to longer linear alkanes. Returning to the SOAP-Ridge MLIPs constructed using total SOAP vectors ($X^{total}$) and far-sighted SOAP vectors ($X^{fs}$) in Fig. \ref{fig:specials_energies}, we also maintain previous trends: intermolecular energy MAEs decrease consistently as the training alkane chain length increases, and MLIPs constructed using $X^{fs}$ consistently outperform MLIPs constructed with $X^{total}$. This re-confirms our earlier claim that re-weighting the feature-space to exclude intramolecular contributions enables higher accuracy when targeting intermolecular energy.

Molecules of different architectures fare less well in extrapolation. For cyclohexane, 4-propylheptane, and 3,3-diethylpentane, we typically converge to higher values (6-10meV/\AA) than within the linear series (1-2meV/\AA). On one hand, this is due to the lack of longer alkyl chains within cyclohexane, 4-propylheptane, and 3,3-diethylpentane -- moving to hexane, heptane, and octane training data is not nearly as beneficial when the longest uninterrupted alkyl chain is three carbons long. For 4-propylheptane and 3,3-diethylpentane, the larger errors are largely attributed to the tertiary and quaternary carbons (Fig.~\ref{fig:kde}), neither of which exists in a linear alkane. Though we see higher errors from the MACE predicted forces for these molecules, we maintain consistent trends in intermolecular energies as predicted by SOAP-Ridge MLIPs with decreasing errors as training alkane chain length increases, and lower errors for all but one far-sighted MLIPs.

Cyclohexane presents a more interesting case. The carbons in cyclohexane have long received different parameterization than those in hexane via popular force fields such as OPLS-AA\cite{kenney_prediction_2016, ghahremanpour_refinement_2022, jorgensen_development_1996}. This has largely been attributed to the constrained torsion angles in cyclohexane, which remove the freely rotating C-C dihedrals that characterize linear alkanes\cite{kenney_prediction_2016}. However, here, the key difference is the local environment distribution: linear alkanes provide very few CH$_2$ environments with neighboring CH$_2$ at the $2-3$\AA~range, whereas cyclohexane contains several such neighbors because of its compact ring geometry. As a result, when the model trained only on octane is evaluated on cyclohexane, it encounters CH$_2$–CH$_2$ configurations that were poorly sampled in the training set, producing the shift observed in Fig.~\ref{fig:kde}.

\section{Conclusions}
In this study, we have demonstrated key behaviors of MLIPs that can be applied to build predictive models for oligomeric systems, with implications on MLIP extrapolation for macromolecular and polymeric materials. 
First, we demonstrate and visualize that, in systems with similar chemistries, extrapolation of forces and energies \emph{is} possible, provided that those environments present in the target molecule are properly sampled within the smaller analogs. Specifically for alkanes, we see a strong reduction in error with butane, and we see the force errors converging once we reach hexane. These results were also demonstrated on longer linear alkanes and different architectures. The latter proved more nuanced and complicated as an extrapolation task, as while non-linear alkanes contain the same chemistry, MLIPs do not consider these aspects explicitly, but rather as emerging from data distributions. And, because the environmental distributions of branched or cyclical alkanes are fundamentally different from linear alkanes, the MLIPs fare worse in extrapolating to these systems. 

Second, we showed that extrapolation of energies results in a learnable shift between different molecules; this can be largely ameliorated by predicting the mean shift in probability distributions as a function of composition. As forces are not affected by constant shifts in energy, the extrapolation of forces between different molecules is straightforward and does not require accounting for this effect.

Finally, we demonstrated a powerful lesson about the hierarchy of energies. Intermolecular contributions are important in determining thermodynamic properties, but have historically been challenging to predict due to their small magnitude relative to the intramolecular component. We have shown that careful partitioning of molecular descriptors can yield MLIPs capable of predicting and extrapolating intermolecular energetics. Given the relative difficulty of computing intra- versus intermolecular energetics, this suggests that the most difficult calculations, those relating the interactions of one molecule to another, are the easiest to translate from one system to another of similar chemistry. 

\section{Research Funding}
This project was funded by the Wisconsin Alumni Research Fund (RKC) and by NSF through the University of Wisconsin Materials Research Science and Engineering Center (DMR-2309000, AL, CC). NH acknowledges support from the Michael and Virginia Conway Professorship. The funders played no role in study design, data collection, analysis and interpretation of data, or the writing of this manuscript. 

\section{Data Availability}
The datasets generated and/or analysed during the current study are available in the Zenodo repository at \textcite{hooven_equilibrium_2025} [DOI: 10.5281/zenodo.17227874]. 

\section{Code Availability}
The underlying code used in this study are available via open-source software through GitHub, namely \texttt{MACE} (\url{https://github.com/ACEsuit/mace}\cite{batatia_mace_2022}), \texttt{scikit-matter} (\url{https://github.com/scikit-learn-contrib/scikit-matter}\cite{goscinski_scikit-matter_2023}), \texttt{featomic} (\url{https://github.com/metatensor/featomic}\cite{bigi_metatensor_2025}), and \texttt{librascal} (\url{https://github.com/lab-cosmo/librascal.git}\cite{musil_efficient_2021}).

\section{Author Contributions}
Project conceptualization (NH, RC), data curation (NH, CC, AL), formal analysis (NH, RC), methodology (NH, RC), supervision (AL, RC), writing (NH, AL, CC, RC).

\section{Competing Interests}
All authors declare no financial or non-financial competing interests. 
\printbibliography

@article{kenney_prediction_2016,
	title = {Prediction of cyclohexane-water distribution coefficients for the {SAMPL5} data set using molecular dynamics simulations with the {OPLS}-{AA} force field},
	volume = {30},
	issn = {0920-654X, 1573-4951},
	url = {http://link.springer.com/10.1007/s10822-016-9949-5},
	doi = {10.1007/s10822-016-9949-5},
	number = {11},
	urldate = {2026-02-18},
	journal = {Journal of Computer-Aided Molecular Design},
	author = {Kenney, Ian M. and Beckstein, Oliver and Iorga, Bogdan I.},
	month = nov,
	year = {2016},
	pages = {1045--1058},
}

@article{jorgensen_interpretable_2026,
	title = {Interpretable visualizations of data spaces for classification problems},
	issn = {2632-2153},
	url = {http://iopscience.iop.org/article/10.1088/2632-2153/ae466e},
	doi = {10.1088/2632-2153/ae466e},
	urldate = {2026-02-17},
	journal = {Machine Learning: Science and Technology},
	author = {Jorgensen, Christian Alexander and Lin, Arthur Y and Vasavada, Rhushil and Cersonsky, Rose},
	year = {2026},
}

@article{goscinski_scikit-matter_2025,
	title = {scikit-matter : {A} {Suite} of {Generalisable} {Machine} {Learning} {Methods} {Born} out of {Chemistry} and {Materials} {Science}},
	volume = {3},
	issn = {2732-5121},
	shorttitle = {scikit-matter},
	url = {https://open-research-europe.ec.europa.eu/articles/3-81/v3},
	doi = {10.12688/openreseurope.15789.3},
	urldate = {2026-02-17},
	journal = {Open Research Europe},
	author = {Goscinski, Alexander and Jorgensen, Christian A. and Principe, Victor Paul and Fraux, Guillaume and Kliavinek, Sergei and Helfrecht, Benjamin Aaron and Vasavada, Rhushil and Loche, Philip and Ceriotti, Michele and Cersonsky, Rose Kathleen},
	month = dec,
	year = {2025},
	pages = {81},
}

@article{jorgensen_potential_2005,
	title = {Potential energy functions for atomic-level simulations of water and organic and biomolecular systems},
	volume = {102},
	issn = {0027-8424, 1091-6490},
	url = {https://pnas.org/doi/full/10.1073/pnas.0408037102},
	doi = {10.1073/pnas.0408037102},
	number = {19},
	urldate = {2026-02-17},
	journal = {Proceedings of the National Academy of Sciences},
	author = {Jorgensen, William L. and Tirado-Rives, Julian},
	month = may,
	year = {2005},
	pages = {6665--6670},
}

@article{ghahremanpour_refinement_2022,
	title = {Refinement of the {OPLS} {Force} {Field} for {Thermodynamics} and {Dynamics} of {Liquid} {Alkanes}},
	volume = {126},
	issn = {1520-6106},
	url = {https://pmc.ncbi.nlm.nih.gov/articles/PMC9939004/},
	doi = {10.1021/acs.jpcb.2c03686},
	number = {31},
	urldate = {2026-02-18},
	journal = {The journal of physical chemistry. B},
	author = {Ghahremanpour, Mohammad M. and Tirado-Rives, Julian and Jorgensen, William L.},
	month = aug,
	year = {2022},
	pages = {5896--5907},
}

@article{chodera_simple_2016,
	title = {A {Simple} {Method} for {Automated} {Equilibration} {Detection} in {Molecular} {Simulations}},
	volume = {12},
	issn = {1549-9618},
	url = {https://doi.org/10.1021/acs.jctc.5b00784},
	doi = {10.1021/acs.jctc.5b00784},
	number = {4},
	urldate = {2026-02-17},
	journal = {Journal of Chemical Theory and Computation},
	publisher = {American Chemical Society},
	author = {Chodera, John D.},
	month = apr,
	year = {2016},
	pages = {1799--1805},
}

@article{dodda_ligpargen_2017,
	title = {{LigParGen} web server: an automatic {OPLS}-{AA} parameter generator for organic ligands},
	volume = {45},
	issn = {0305-1048},
	shorttitle = {{LigParGen} web server},
	url = {https://doi.org/10.1093/nar/gkx312},
	doi = {10.1093/nar/gkx312},
	number = {W1},
	urldate = {2026-02-17},
	journal = {Nucleic Acids Research},
	author = {Dodda, Leela S. and Cabeza de Vaca, Israel and Tirado-Rives, Julian and Jorgensen, William L.},
	month = jul,
	year = {2017},
	pages = {W331--W336},
}

@article{dodda_114cm1a-lbcc_2017,
	title = {1.14*{CM1A}-{LBCC}: {Localized} {Bond}-{Charge} {Corrected} {CM1A} {Charges} for {Condensed}-{Phase} {Simulations}},
	volume = {121},
	issn = {1520-6106},
	shorttitle = {1.14*{CM1A}-{LBCC}},
	url = {https://doi.org/10.1021/acs.jpcb.7b00272},
	doi = {10.1021/acs.jpcb.7b00272},
	number = {15},
	urldate = {2026-02-17},
	journal = {The Journal of Physical Chemistry B},
	publisher = {American Chemical Society},
	author = {Dodda, Leela S. and Vilseck, Jonah Z. and Tirado-Rives, Julian and Jorgensen, William L.},
	month = apr,
	year = {2017},
	pages = {3864--3870},
}

@article{jorgensen_development_1996,
	title = {Development and {Testing} of the {OPLS} {All}-{Atom} {Force} {Field} on {Conformational} {Energetics} and {Properties} of {Organic} {Liquids}},
	volume = {118},
	issn = {0002-7863},
	url = {https://doi.org/10.1021/ja9621760},
	doi = {10.1021/ja9621760},
	number = {45},
	urldate = {2026-02-17},
	journal = {Journal of the American Chemical Society},
	publisher = {American Chemical Society},
	author = {Jorgensen, William L. and Maxwell, David S. and Tirado-Rives, Julian},
	month = nov,
	year = {1996},
	pages = {11225--11236},
}

@misc{hooven_equilibrium_2025,
	address = {10.5281/zenodo.17227874},
	title = {Equilibrium {Sampling} of {N}-{Alkanes} in {Liquid} {Phase} at {Identical} {Pressure} and {Temperature}},
	url = {doi.org/10.5281/zenodo.17227874},
	doi = {10.5281/zenodo.17227874},
	publisher = {Zenodo},
	author = {Hooven, Natalie E and Lin, Arthur Y. and Cersonsky, Rose K.},
	year = {2025},
	doi = {10.5281/zenodo.17227874},
}

@article{elstner_self-consistent-charge_1998,
	title = {Self-consistent-charge density-functional tight-binding method for simulations of complex materials properties},
	volume = {58},
	copyright = {http://link.aps.org/licenses/aps-default-license},
	issn = {0163-1829, 1095-3795},
	url = {https://link.aps.org/doi/10.1103/PhysRevB.58.7260},
	doi = {10.1103/PhysRevB.58.7260},
	number = {11},
	urldate = {2025-02-02},
	journal = {Physical Review B},
	author = {Elstner, M. and Porezag, D. and Jungnickel, G. and Elsner, J. and Haugk, M. and Frauenheim, Th. and Suhai, S. and Seifert, G.},
	month = sep,
	year = {1998},
	pages = {7260--7268},
}

@article{gordon_fragmentation_2012,
	title = {Fragmentation {Methods}: {A} {Route} to {Accurate} {Calculations} on {Large} {Systems}},
	volume = {112},
	issn = {0009-2665, 1520-6890},
	shorttitle = {Fragmentation {Methods}},
	url = {https://pubs.acs.org/doi/10.1021/cr200093j},
	doi = {10.1021/cr200093j},
	number = {1},
	urldate = {2025-09-29},
	journal = {Chemical Reviews},
	author = {Gordon, Mark S. and Fedorov, Dmitri G. and Pruitt, Spencer R. and Slipchenko, Lyudmila V.},
	month = jan,
	year = {2012},
	pages = {632--672},
}

@article{thompson_lammps_2022,
	title = {{LAMMPS} - a flexible simulation tool for particle-based materials modeling at the atomic, meso, and continuum scales},
	volume = {271},
	issn = {00104655},
	url = {https://linkinghub.elsevier.com/retrieve/pii/S0010465521002836},
	doi = {10.1016/j.cpc.2021.108171},
	urldate = {2025-09-29},
	journal = {Computer Physics Communications},
	author = {Thompson, Aidan P. and Aktulga, H. Metin and Berger, Richard and Bolintineanu, Dan S. and Brown, W. Michael and Crozier, Paul S. and In 'T Veld, Pieter J. and Kohlmeyer, Axel and Moore, Stan G. and Nguyen, Trung Dac and Shan, Ray and Stevens, Mark J. and Tranchida, Julien and Trott, Christian and Plimpton, Steven J.},
	month = feb,
	year = {2022},
	pages = {108171},
}

@article{zuo_performance_2020,
	title = {Performance and {Cost} {Assessment} of {Machine} {Learning} {Interatomic} {Potentials}},
	volume = {124},
	copyright = {https://doi.org/10.15223/policy-029},
	issn = {1089-5639, 1520-5215},
	url = {https://pubs.acs.org/doi/10.1021/acs.jpca.9b08723},
	doi = {10.1021/acs.jpca.9b08723},
	number = {4},
	urldate = {2025-09-29},
	journal = {The Journal of Physical Chemistry A},
	author = {Zuo, Yunxing and Chen, Chi and Li, Xiangguo and Deng, Zhi and Chen, Yiming and Behler, Jörg and Csányi, Gábor and Shapeev, Alexander V. and Thompson, Aidan P. and Wood, Mitchell A. and Ong, Shyue Ping},
	month = jan,
	year = {2020},
	pages = {731--745},
}

@article{unke_biomolecular_2024,
	title = {Biomolecular dynamics with machine-learned quantum-mechanical force fields trained on diverse chemical fragments},
	volume = {10},
	issn = {2375-2548},
	url = {https://www.science.org/doi/10.1126/sciadv.adn4397},
	doi = {10.1126/sciadv.adn4397},
	number = {14},
	urldate = {2025-09-29},
	journal = {Science Advances},
	author = {Unke, Oliver T. and Stöhr, Martin and Ganscha, Stefan and Unterthiner, Thomas and Maennel, Hartmut and Kashubin, Sergii and Ahlin, Daniel and Gastegger, Michael and Medrano Sandonas, Leonardo and Berryman, Joshua T. and Tkatchenko, Alexandre and Müller, Klaus-Robert},
	month = apr,
	year = {2024},
	pages = {eadn4397},
}

@article{musil_efficient_2021,
	title = {Efficient implementation of atom-density representations},
	volume = {154},
	issn = {0021-9606, 1089-7690},
	url = {https://pubs.aip.org/jcp/article/154/11/114109/315400/Efficient-implementation-of-atom-density},
	doi = {10.1063/5.0044689},
	number = {11},
	urldate = {2025-09-29},
	journal = {The Journal of Chemical Physics},
	author = {Musil, Félix and Veit, Max and Goscinski, Alexander and Fraux, Guillaume and Willatt, Michael J. and Stricker, Markus and Junge, Till and Ceriotti, Michele},
	month = mar,
	year = {2021},
	pages = {114109},
}

@article{parker_levels_2014,
	title = {Levels of symmetry adapted perturbation theory ({SAPT}). {I}. {Efficiency} and performance for interaction energies},
	volume = {140},
	issn = {0021-9606, 1089-7690},
	url = {https://pubs.aip.org/jcp/article/140/9/094106/193599/Levels-of-symmetry-adapted-perturbation-theory},
	doi = {10.1063/1.4867135},
	number = {9},
	urldate = {2025-09-29},
	journal = {The Journal of Chemical Physics},
	author = {Parker, Trent M. and Burns, Lori A. and Parrish, Robert M. and Ryno, Alden G. and Sherrill, C. David},
	month = mar,
	year = {2014},
	pages = {094106},
}

@article{mohanty_development_2023,
	title = {Development of scalable and generalizable machine learned force field for polymers},
	volume = {13},
	issn = {2045-2322},
	url = {https://www.nature.com/articles/s41598-023-43804-5},
	doi = {10.1038/s41598-023-43804-5},
	number = {1},
	urldate = {2025-09-29},
	journal = {Scientific Reports},
	author = {Mohanty, Shaswat and Stevenson, James and Browning, Andrea R. and Jacobson, Leif and Leswing, Karl and Halls, Mathew D. and Afzal, Mohammad Atif Faiz},
	month = oct,
	year = {2023},
	pages = {17251},
}

@article{dral_hierarchical_2020,
	title = {Hierarchical machine learning of potential energy surfaces},
	volume = {152},
	issn = {0021-9606, 1089-7690},
	url = {https://pubs.aip.org/jcp/article/152/20/204110/1059666/Hierarchical-machine-learning-of-potential-energy},
	doi = {10.1063/5.0006498},
	number = {20},
	urldate = {2025-09-29},
	journal = {The Journal of Chemical Physics},
	author = {Dral, Pavlo O. and Owens, Alec and Dral, Alexey and Csányi, Gábor},
	month = may,
	year = {2020},
	pages = {204110},
}

@article{jacobson_transferable_2022,
	title = {Transferable {Neural} {Network} {Potential} {Energy} {Surfaces} for {Closed}-{Shell} {Organic} {Molecules}: {Extension} to {Ions}},
	volume = {18},
	copyright = {https://doi.org/10.15223/policy-029},
	issn = {1549-9618, 1549-9626},
	shorttitle = {Transferable {Neural} {Network} {Potential} {Energy} {Surfaces} for {Closed}-{Shell} {Organic} {Molecules}},
	url = {https://pubs.acs.org/doi/10.1021/acs.jctc.1c00821},
	doi = {10.1021/acs.jctc.1c00821},
	number = {4},
	urldate = {2025-09-29},
	journal = {Journal of Chemical Theory and Computation},
	author = {Jacobson, Leif D. and Stevenson, James M. and Ramezanghorbani, Farhad and Ghoreishi, Delaram and Leswing, Karl and Harder, Edward D. and Abel, Robert},
	month = apr,
	year = {2022},
	pages = {2354--2366},
}

@article{batzner_e3-equivariant_2022,
	title = {E(3)-equivariant graph neural networks for data-efficient and accurate interatomic potentials},
	volume = {13},
	issn = {2041-1723},
	url = {https://www.nature.com/articles/s41467-022-29939-5},
	doi = {10.1038/s41467-022-29939-5},
	number = {1},
	urldate = {2025-09-29},
	journal = {Nature Communications},
	author = {Batzner, Simon and Musaelian, Albert and Sun, Lixin and Geiger, Mario and Mailoa, Jonathan P. and Kornbluth, Mordechai and Molinari, Nicola and Smidt, Tess E. and Kozinsky, Boris},
	month = may,
	year = {2022},
	pages = {2453},
}

@misc{greg_landrum_rdkitrdkit_2025,
	title = {rdkit/rdkit: 2025\_03\_5},
	copyright = {BSD 3-Clause "New" or "Revised" License},
	shorttitle = {rdkit/rdkit},
	url = {https://zenodo.org/doi/10.5281/zenodo.16439048},
	urldate = {2025-08-26},
	publisher = {Zenodo},
	author = {Greg Landrum and Paolo Tosco and Brian Kelley and Ricardo Rodriguez and David Cosgrove and Riccardo Vianello and sriniker and Peter Gedeck and Gareth Jones and Eisuke Kawashima and NadineSchneider and Dan Nealschneider and Andrew Dalke and tadhurst-cdd and Matt Swain and Brian Cole and Samo Turk and Aleksandr Savelev and Alain Vaucher and Maciej Wójcikowski and Ichiru Take and Hussein Faara and Vincent F. Scalfani and Rachel Walker and Daniel Probst and Kazuya Ujihara and Niels Maeder and Jeremy Monat and Juuso Lehtivarjo and Godan, Guillaume},
	month = jul,
	year = {2025},
}

@inproceedings{ramasubramani_signac_2018,
	title = {signac: {A} {Python} framework for data and workflow management},
	shorttitle = {signac},
	url = {https://doi.curvenote.com/10.25080/Majora-4af1f417-016},
	doi = {10.25080/Majora-4af1f417-016},
	urldate = {2025-08-26},
	author = {Ramasubramani, Vyas and Adorf, Carl and Dodd, Paul and Dice, Bradley and Glotzer, Sharon},
	year = {2018},
	pages = {152--159},
}

@article{cheng_cartesian_2024,
	title = {Cartesian atomic cluster expansion for machine learning interatomic potentials},
	volume = {10},
	issn = {2057-3960},
	url = {https://www.nature.com/articles/s41524-024-01332-4},
	doi = {10.1038/s41524-024-01332-4},
	number = {1},
	urldate = {2025-09-22},
	journal = {npj Computational Materials},
	author = {Cheng, Bingqing},
	month = jul,
	year = {2024},
	pages = {157},
}

@article{behler_generalized_2007,
	title = {Generalized {Neural}-{Network} {Representation} of {High}-{Dimensional} {Potential}-{Energy} {Surfaces}},
	volume = {98},
	copyright = {http://link.aps.org/licenses/aps-default-license},
	issn = {0031-9007, 1079-7114},
	url = {https://link.aps.org/doi/10.1103/PhysRevLett.98.146401},
	doi = {10.1103/PhysRevLett.98.146401},
	number = {14},
	urldate = {2025-09-22},
	journal = {Physical Review Letters},
	author = {Behler, Jörg and Parrinello, Michele},
	month = apr,
	year = {2007},
	pages = {146401},
}

@article{xu_new_2022,
	title = {New {Opportunity}: {Machine} {Learning} for {Polymer} {Materials} {Design} and {Discovery}},
	volume = {5},
	copyright = {© 2022 Wiley-VCH GmbH},
	issn = {2513-0390},
	shorttitle = {New {Opportunity}},
	url = {https://onlinelibrary.wiley.com/doi/abs/10.1002/adts.202100565},
	doi = {10.1002/adts.202100565},
	number = {5},
	urldate = {2025-09-19},
	journal = {Advanced Theory and Simulations},
	author = {Xu, Pengcheng and Chen, Huimin and Li, Minjie and Lu, Wencong},
	year = {2022},
	pages = {2100565},
}

@misc{bigi_metatensor_2025,
	title = {Metatensor and metatomic: foundational libraries for interoperable atomistic machine learning},
	shorttitle = {Metatensor and metatomic},
	url = {http://arxiv.org/abs/2508.15704},
	doi = {10.48550/arXiv.2508.15704},
	urldate = {2025-09-19},
	publisher = {arXiv},
	author = {Bigi, Filippo and Abbott, Joseph W. and Loche, Philip and Mazitov, Arslan and Tisi, Davide and Langer, Marcel F. and Goscinski, Alexander and Pegolo, Paolo and Chong, Sanggyu and Goswami, Rohit and Chorna, Sofiia and Kellner, Matthias and Ceriotti, Michele and Fraux, Guillaume},
	month = aug,
	year = {2025},
	note = {arXiv:2508.15704 [physics]},
}

@article{lindsey_chimes_2017,
	title = {{ChIMES}: {A} {Force} {Matched} {Potential} with {Explicit} {Three}-{Body} {Interactions} for {Molten} {Carbon}},
	volume = {13},
	issn = {1549-9618, 1549-9626},
	shorttitle = {{ChIMES}},
	url = {https://pubs.acs.org/doi/10.1021/acs.jctc.7b00867},
	doi = {10.1021/acs.jctc.7b00867},
	number = {12},
	urldate = {2025-02-02},
	journal = {Journal of Chemical Theory and Computation},
	author = {Lindsey, Rebecca K. and Fried, Laurence E. and Goldman, Nir},
	month = dec,
	year = {2017},
	pages = {6222--6229},
}

@article{imbalzano_automatic_2018,
	title = {Automatic selection of atomic fingerprints and reference configurations for machine-learning potentials},
	volume = {148},
	issn = {0021-9606, 1089-7690},
	url = {https://pubs.aip.org/jcp/article/148/24/241730/962573/Automatic-selection-of-atomic-fingerprints-and},
	doi = {10.1063/1.5024611},
	number = {24},
	urldate = {2025-08-26},
	journal = {The Journal of Chemical Physics},
	author = {Imbalzano, Giulio and Anelli, Andrea and Giofré, Daniele and Klees, Sinja and Behler, Jörg and Ceriotti, Michele},
	month = jun,
	year = {2018},
	pages = {241730},
}

@article{grisafi_transferable_2019,
	title = {Transferable {Machine}-{Learning} {Model} of the {Electron} {Density}},
	volume = {5},
	copyright = {http://pubs.acs.org/page/policy/authorchoice\_termsofuse.html},
	issn = {2374-7943, 2374-7951},
	url = {https://pubs.acs.org/doi/10.1021/acscentsci.8b00551},
	doi = {10.1021/acscentsci.8b00551},
	number = {1},
	urldate = {2025-02-02},
	journal = {ACS Central Science},
	author = {Grisafi, Andrea and Fabrizio, Alberto and Meyer, Benjamin and Wilkins, David M. and Corminboeuf, Clemence and Ceriotti, Michele},
	month = jan,
	year = {2019},
	pages = {57--64},
}

@article{goscinski_scikit-matter_2023,
	title = {scikit-matter : {A} {Suite} of {Generalisable} {Machine} {Learning} {Methods} {Born} out of {Chemistry} and {Materials} {Science}},
	volume = {3},
	issn = {2732-5121},
	shorttitle = {scikit-matter},
	url = {https://open-research-europe.ec.europa.eu/articles/3-81/v2},
	doi = {10.12688/openreseurope.15789.2},
	urldate = {2025-02-02},
	journal = {Open Research Europe},
	author = {Goscinski, Alexander and Principe, Victor Paul and Fraux, Guillaume and Kliavinek, Sergei and Helfrecht, Benjamin Aaron and Loche, Philip and Ceriotti, Michele and Cersonsky, Rose Kathleen},
	month = sep,
	year = {2023},
	pages = {81},
}

@article{drautz_atomic_2019,
	title = {Atomic cluster expansion for accurate and transferable interatomic potentials},
	volume = {99},
	issn = {2469-9950, 2469-9969},
	url = {https://link.aps.org/doi/10.1103/PhysRevB.99.014104},
	doi = {10.1103/PhysRevB.99.014104},
	number = {1},
	urldate = {2025-02-02},
	journal = {Physical Review B},
	author = {Drautz, Ralf},
	month = jan,
	year = {2019},
	pages = {014104},
}

@article{cheng_ab_2019,
	title = {Ab initio thermodynamics of liquid and solid water},
	volume = {116},
	issn = {0027-8424, 1091-6490},
	url = {https://pnas.org/doi/full/10.1073/pnas.1815117116},
	doi = {10.1073/pnas.1815117116},
	number = {4},
	urldate = {2025-09-19},
	journal = {Proceedings of the National Academy of Sciences},
	author = {Cheng, Bingqing and Engel, Edgar A. and Behler, Jörg and Dellago, Christoph and Ceriotti, Michele},
	month = jan,
	year = {2019},
	pages = {1110--1115},
}

@article{winter_simulations_2023,
	title = {Simulations with machine learning potentials identify the ion conduction mechanism mediating non-{Arrhenius} behavior in {LGPS}},
	volume = {5},
	issn = {2515-7655},
	url = {https://dx.doi.org/10.1088/2515-7655/acbbef},
	doi = {10.1088/2515-7655/acbbef},
	number = {2},
	urldate = {2025-09-19},
	journal = {Journal of Physics: Energy},
	publisher = {IOP Publishing},
	author = {Winter, Gavin and Gómez-Bombarelli, Rafael},
	month = mar,
	year = {2023},
	pages = {024004},
}

@article{radova_fine-tuning_2025,
	title = {Fine-tuning foundation models of materials interatomic potentials with frozen transfer learning},
	volume = {11},
	copyright = {2025 The Author(s)},
	issn = {2057-3960},
	url = {https://www.nature.com/articles/s41524-025-01727-x},
	doi = {10.1038/s41524-025-01727-x},
	number = {1},
	urldate = {2025-09-19},
	journal = {npj Computational Materials},
	publisher = {Nature Publishing Group},
	author = {Radova, Mariia and Stark, Wojciech G. and Allen, Connor S. and Maurer, Reinhard J. and Bartók, Albert P.},
	month = jul,
	year = {2025},
	pages = {237},
}

@article{bartok_representing_2013,
	title = {On representing chemical environments},
	volume = {87},
	copyright = {http://link.aps.org/licenses/aps-default-license},
	issn = {1098-0121, 1550-235X},
	url = {https://link.aps.org/doi/10.1103/PhysRevB.87.184115},
	doi = {10.1103/PhysRevB.87.184115},
	number = {18},
	urldate = {2025-01-29},
	journal = {Physical Review B},
	author = {Bartók, Albert P. and Kondor, Risi and Csányi, Gábor},
	month = may,
	year = {2013},
	pages = {184115},
}

@article{goodwin_transferability_2024,
	title = {Transferability and {Accuracy} of {Ionic} {Liquid} {Simulations} with {Equivariant} {Machine} {Learning} {Interatomic} {Potentials}},
	volume = {15},
	url = {https://doi.org/10.1021/acs.jpclett.4c01942},
	doi = {10.1021/acs.jpclett.4c01942},
	number = {30},
	urldate = {2025-09-19},
	journal = {The Journal of Physical Chemistry Letters},
	publisher = {American Chemical Society},
	author = {Goodwin, Zachary A. H. and Wenny, Malia B. and Yang, Julia H. and Cepellotti, Andrea and Ding, Jingxuan and Bystrom, Kyle and Duschatko, Blake R. and Johansson, Anders and Sun, Lixin and Batzner, Simon and Musaelian, Albert and Mason, Jarad A. and Kozinsky, Boris and Molinari, Nicola},
	month = aug,
	year = {2024},
	pages = {7539--7547},
}

@article{zhang_phase_2021,
	title = {Phase {Diagram} of a {Deep} {Potential} {Water} {Model}},
	volume = {126},
	url = {https://link.aps.org/doi/10.1103/PhysRevLett.126.236001},
	doi = {10.1103/PhysRevLett.126.236001},
	number = {23},
	urldate = {2025-09-19},
	journal = {Physical Review Letters},
	publisher = {American Physical Society},
	author = {Zhang, Linfeng and Wang, Han and Car, Roberto and E, Weinan},
	month = jun,
	year = {2021},
	pages = {236001},
}

@misc{yuan_foundation_2025,
	title = {Foundation {Models} for {Atomistic} {Simulation} of {Chemistry} and {Materials}},
	url = {http://arxiv.org/abs/2503.10538},
	doi = {10.48550/arXiv.2503.10538},
	urldate = {2025-09-19},
	publisher = {arXiv},
	author = {Yuan, Eric C.-Y. and Liu, Yunsheng and Chen, Junmin and Zhong, Peichen and Raja, Sanjeev and Kreiman, Tobias and Vargas, Santiago and Xu, Wenbin and Head-Gordon, Martin and Yang, Chao and Blau, Samuel M. and Cheng, Bingqing and Krishnapriyan, Aditi and Head-Gordon, Teresa},
	month = jun,
	year = {2025},
	note = {arXiv:2503.10538 [physics]},
}

@article{groom_cambridge_2016,
	title = {The {Cambridge} {Structural} {Database}},
	volume = {72},
	issn = {2052-5206},
	url = {https://journals.iucr.org/paper?S2052520616003954},
	doi = {10.1107/S2052520616003954},
	number = {2},
	urldate = {2025-09-09},
	journal = {Acta Crystallographica Section B Structural Science, Crystal Engineering and Materials},
	author = {Groom, Colin R. and Bruno, Ian J. and Lightfoot, Matthew P. and Ward, Suzanna C.},
	month = apr,
	year = {2016},
	pages = {171--179},
}

@misc{linstrom_nist_1997,
	title = {{NIST} {Chemistry} {WebBook}, {NIST} {Standard} {Reference} {Database} 69},
	copyright = {License Information for NIST data},
	url = {http://webbook.nist.gov/chemistry/},
	doi = {10.18434/T4D303},
	urldate = {2025-08-26},
	publisher = {National Institute of Standards and Technology},
	author = {Linstrom, Peter},
	year = {1997},
}

@article{pedregosa_scikit-learn_2011,
	title = {Scikit-learn: {Machine} {Learning} in {Python}},
	volume = {12},
	url = {http://jmlr.org/papers/v12/pedregosa11a.html},
	number = {85},
	journal = {Journal of Machine Learning Research},
	author = {Pedregosa, Fabian and Varoquaux, Gael and Gramfort, Alexandre and Michel, Vincent and Thirion, Bertrand and Grisel, Olivier and Blondel, Mathieu and Prettenhofer, Peter and Weiss, Ron and Dubourg, Vincent and Vanderplas, Jake and Passos, Alexandre and Cournapeau, David and Brucher, Matthieu and Perrot, Matthieu and Duchesnay, Edouard},
	year = {2011},
	pages = {2825--2830},
}

@misc{jorgensen_interpretable_2025,
	title = {Interpretable {Visualizations} of {Data} {Spaces} for {Classification} {Problems}},
	copyright = {Creative Commons Attribution 4.0 International},
	url = {https://arxiv.org/abs/2503.05861},
	doi = {10.48550/ARXIV.2503.05861},
	urldate = {2025-09-09},
	publisher = {arXiv},
	author = {Jorgensen, Christian and Lin, Arthur Y. and Vasavada, Rhushil and Cersonsky, Rose K.},
	year = {2025},
	note = {Version Number: 2},
}

@misc{mazitov_massive_2025,
	title = {Massive {Atomic} {Diversity}: a compact universal dataset for atomistic machine learning},
	shorttitle = {Massive {Atomic} {Diversity}},
	url = {http://arxiv.org/abs/2506.19674},
	doi = {10.48550/arXiv.2506.19674},
	urldate = {2025-06-27},
	publisher = {arXiv},
	author = {Mazitov, Arslan and Chorna, Sofiia and Fraux, Guillaume and Bercx, Marnik and Pizzi, Giovanni and De, Sandip and Ceriotti, Michele},
	month = jun,
	year = {2025},
	note = {arXiv:2506.19674 [cond-mat]},
}

@misc{batatia_foundation_2024,
	title = {A foundation model for atomistic materials chemistry},
	url = {http://arxiv.org/abs/2401.00096},
	doi = {10.48550/arXiv.2401.00096},
	urldate = {2025-06-09},
	publisher = {arXiv},
	author = {Batatia, Ilyes and Benner, Philipp and Chiang, Yuan and Elena, Alin M. and Kovacs, David P. and Riebesell, Janosh and Advincula, Xavier R. and Asta, Mark and Avaylon, Matthew and Baldwin, William J. and Berger, Fabian and Bernstein, Noam and Bhowmik, Arghya and Blau, Samuel M. and Carare, Vlad and Darby, James P. and De, Sandip and Pia, Flaviano Della and Deringer, Volker L. and Elijosius, Rokas and El-Machachi, Zakariya and Falcioni, Fabio and Fako, Edvin and Ferrari, Andrea C. and Genreith-Schriever, Annalena and George, Janine and Goodall, Rhys E. A. and Grey, Clare P. and Grigorev, Petr and Han, Shuang and Handley, Will and Heenen, Hendrik H. and Hermansson, Kersti and Holm, Christian and Jaafar, Jad and Hofmann, Stephan and Jakob, Konstantin S. and Jung, Hyunwook and Kapil, Venkat and Kaplan, Aaron D. and Karimitari, Nima and Kermode, James R. and Kroupa, Namu and Kullgren, Jolla and Kuner, Matthew C. and Kuryla, Domantas and Liepuoniute, Guoda and Margraf, Johannes T. and Magdau, Ioan-Bogdan and Michaelides, Angelos and Moore, J. Harry and Naik, Aakash A. and Niblett, Samuel P. and Norwood, Sam Walton and O'Neill, Niamh and Ortner, Christoph and Persson, Kristin A. and Reuter, Karsten and Rosen, Andrew S. and Schaaf, Lars L. and Schran, Christoph and Shi, Benjamin X. and Sivonxay, Eric and Stenczel, Tamas K and Svahn, Viktor and Sutton, Christopher and Swinburne, Thomas D. and Tilly, Jules and Oord, Cas van der and Varga-Umbrich, Eszter and Vegge, Tejs and Vondrak, Martin and Wang, Yangshuai and Witt, William C. and Zills, Fabian and Csanyi, Gabor},
	month = mar,
	year = {2024},
	note = {arXiv:2401.00096 [physics]},
}

@article{adorf_simple_2018,
	title = {Simple data and workflow management with the signac framework},
	volume = {146},
	issn = {0927-0256},
	url = {https://www.sciencedirect.com/science/article/pii/S0927025618300429},
	doi = {10.1016/j.commatsci.2018.01.035},
	urldate = {2025-08-26},
	journal = {Computational Materials Science},
	author = {Adorf, Carl S. and Dodd, Paul M. and Ramasubramani, Vyas and Glotzer, Sharon C.},
	month = apr,
	year = {2018},
	pages = {220--229},
}

@article{lin_anisoap_2025,
	title = {{AniSOAP}: {Machine} {Learning} {Representations} for {Coarse}-grained and {Non}-spherical {Systems}},
	shorttitle = {{AniSOAP}},
	url = {https://joss.theoj.org/papers/4f031c830d4790cce21dd630588db665},
	urldate = {2025-01-16},
	journal = {Journal of Open Source Software},
	author = {Lin, Arthur Y. and Ortengren, Lucas and Hwang, Seonwoo and Cho, Yong-Cheol and Nigam, Jigyasa and Cersonsky, Rose K.},
	year = {2025},
}

@article{lin_expanding_2024,
	title = {Expanding density-correlation machine learning representations for anisotropic coarse-grained particles},
	volume = {161},
	issn = {0021-9606},
	url = {https://doi.org/10.1063/5.0210910},
	doi = {10.1063/5.0210910},
	number = {7},
	urldate = {2024-08-21},
	journal = {The Journal of Chemical Physics},
	author = {Lin, Arthur and Huguenin-Dumittan, Kevin K. and Cho, Yong-Cheol and Nigam, Jigyasa and Cersonsky, Rose K.},
	month = aug,
	year = {2024},
	pages = {074112},
}

@article{helfrecht_structure-property_2020,
	title = {Structure-property maps with kernel principal covariates regression},
	volume = {1},
	issn = {2632-2153},
	url = {https://iopscience.iop.org/article/10.1088/2632-2153/aba9ef},
	doi = {10.1088/2632-2153/aba9ef},
	number = {4},
	urldate = {2023-05-23},
	journal = {Machine Learning: Science and Technology},
	author = {Helfrecht, Benjamin A and Cersonsky, Rose K and Fraux, Guillaume and Ceriotti, Michele},
	month = oct,
	year = {2020},
	pages = {045021},
}

@article{cersonsky_data-driven_2023,
	title = {A data-driven interpretation of the stability of organic molecular crystals},
	volume = {14},
	issn = {2041-6520, 2041-6539},
	url = {http://xlink.rsc.org/?DOI=D2SC06198H},
	doi = {10.1039/D2SC06198H},
	number = {5},
	urldate = {2023-05-23},
	journal = {Chemical Science},
	author = {Cersonsky, Rose K. and Pakhnova, Maria and Engel, Edgar A. and Ceriotti, Michele},
	month = feb,
	year = {2023},
	pages = {1272--1285},
}

@article{cersonsky_improving_2021,
	title = {Improving sample and feature selection with principal covariates regression},
	volume = {2},
	issn = {2632-2153},
	url = {https://iopscience.iop.org/article/10.1088/2632-2153/abfe7c},
	doi = {10.1088/2632-2153/abfe7c},
	number = {3},
	urldate = {2023-05-23},
	journal = {Machine Learning: Science and Technology},
	author = {Cersonsky, Rose K and Helfrecht, Benjamin A and Engel, Edgar A and Kliavinek, Sergei and Ceriotti, Michele},
	month = may,
	year = {2021},
	pages = {035038},
}

@article{han_deep_2018,
	title = {Deep {Potential}: {A} {General} {Representation} of a {Many}-{Body} {Potential} {Energy} {Surface}},
	volume = {23},
	copyright = {https://creativecommons.org/licenses/by/4.0/},
	issn = {1991-7120, 1815-2406},
	shorttitle = {Deep {Potential}},
	url = {https://global-sci.com/article/80035/deep-potential-a-general-representation-of-a-many-body-potential-energy-surface},
	doi = {10.4208/cicp.OA-2017-0213},
	number = {3},
	urldate = {2025-02-02},
	journal = {Communications in Computational Physics},
	author = {Han, Jiequn and Zhang, Linfeng and Car, Roberto and Weinan, E},
	month = jan,
	year = {2018},
	pages = {629--639},
}

@misc{batatia_mace_2022,
	title = {{MACE}: {Higher} {Order} {Equivariant} {Message} {Passing} {Neural} {Networks} for {Fast} and {Accurate} {Force} {Fields}},
	copyright = {arXiv.org perpetual, non-exclusive license},
	shorttitle = {{MACE}},
	url = {https://arxiv.org/abs/2206.07697},
	doi = {10.48550/ARXIV.2206.07697},
	urldate = {2025-02-02},
	publisher = {arXiv},
	author = {Batatia, Ilyes and Kovács, Dávid Péter and Simm, Gregor N. C. and Ortner, Christoph and Csányi, Gábor},
	year = {2022},
	note = {Version Number: 2},
}

\clearpage
\appendix
\renewcommand{\thefigure}{S\arabic{figure}}
\setcounter{figure}{0} 

\section{Additional methodological details}
\label{si:methods}

MACE MLIPs were constructed using the following command and MACE (v.0.3.13):

\begin{lstlisting}[language=bash,caption={Training script for MACE MLIPs}]

raw_dir=$(python -c "import mace; print(mace.__file__)"); dir=${raw_dir:0:-12}
python $dir/cli/run_train.py \
    --name="MACE_model" \
    --train_file="$1" \
    --valid_fraction=0.05 \
    --config_type_weights='{"Default":1.0}' \
    --E0s='{1:-6.492647589968434, 6:-38.054950840332474}' \
    --model="MACE" \
    --hidden_irreps='64x0e + 64x1o' \
    --r_max=5.0 \
    --batch_size=1 \
    --max_num_epochs=1500 \
    --energy_key="energy_eV" \
    --forces_key="forces" \
    --charges_key="charges" \
    --amsgrad \
    --restart_latest \
    --device=cuda \
    --patience=20 \
    --default_dtype="float32" \
\end{lstlisting}

SOAP vectors for SOAP-Ridge MLIPs were computed using the following hyperparameters with librascal (v.0.0.1):
\begin{lstlisting}[language=bash,caption={Hyperparameters of SOAP vectors for SOAP-Ridge MLIPs}]
{
    "interaction_cutoff": 7,
    "max_radial": 8,
    "max_angular": 4,
    "gaussian_sigma_constant": 0.3,
    "gaussian_sigma_type": "Constant",
    "cutoff_smooth_width": 0.5,
    "radial_basis": "GTO",
    "cutoff_function_type": "RadialScaling",
    "cutoff_function_parameters": {
        "rate": 1.5,
        "exponent": 3.0,
        "scale": 2.0
    }
}
\end{lstlisting}

\clearpage
\section{Supporting figures to the text}

\begin{figure*}[ht!]
    \begin{subfigure}[t]{0.45\linewidth}
    \includegraphics[width=\linewidth]{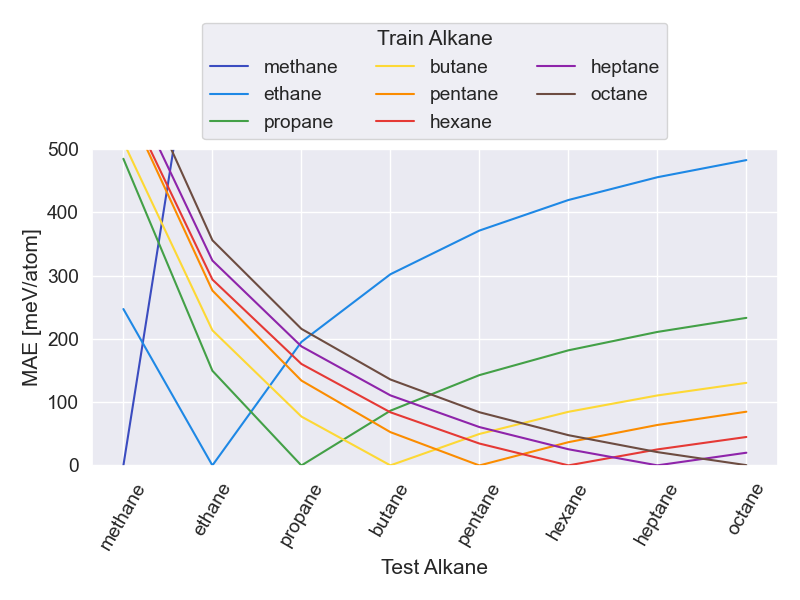}
    \caption{Alternative representation of Fig.~\ref{fig:energies}.}
    \label{si:energies}
    \end{subfigure}
    \begin{subfigure}[t]{0.45\linewidth}
    \includegraphics[width=\linewidth]{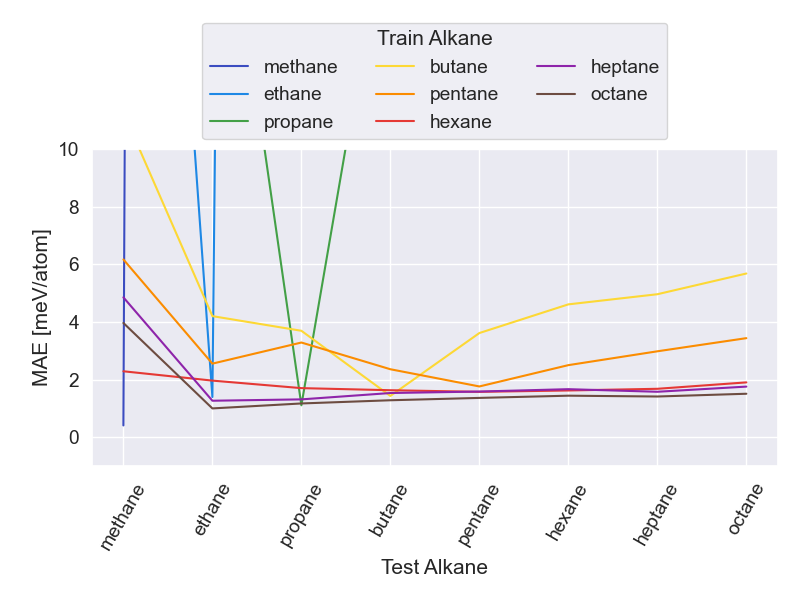}
    \caption{Alternative representation of Fig.~\ref{fig:forces}.}
    \label{si:forces}
    \end{subfigure}
    \caption{Additional visualizations of MACE models to predict total energy and forces.}
\end{figure*}

\begin{figure}[ht!]
    \includegraphics[width=0.4\linewidth]{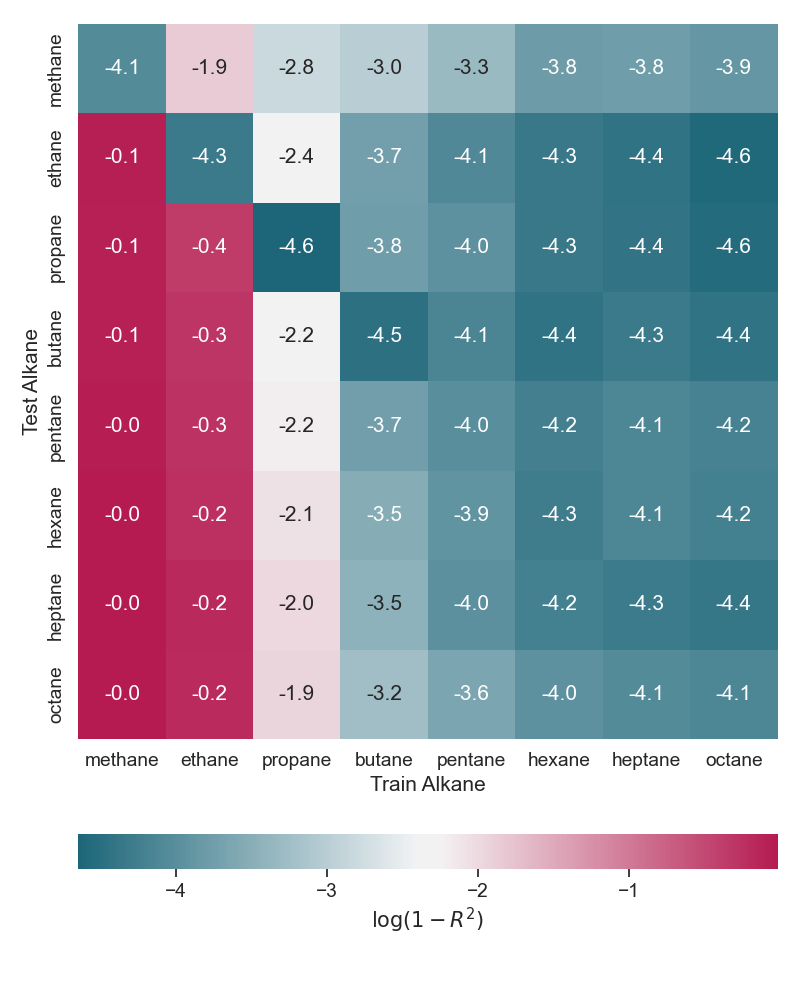}
    \caption{Map of the coefficient of determination between true and predicted values (represented as the logarithm of $1-R^2$)}
    \label{si:colinear}
\end{figure}

\begin{figure*}[ht!]
    \begin{subfigure}[t]{0.45\linewidth}
    \includegraphics[width=0.9\linewidth]{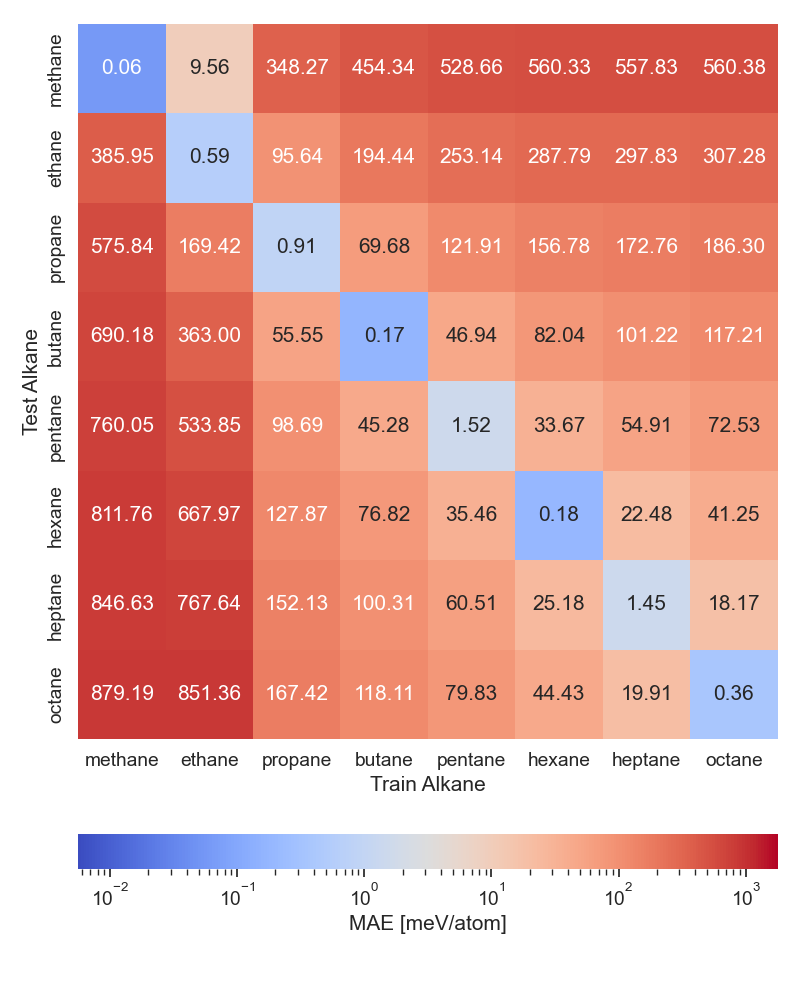}
    \caption{Energy-fitting results of MACE MLIP for different train (x-axis) and test (y-axis) pairings, respectively. Color indicates the mean-absolute-error (MAE) in units of meV/atom, respectively.}
    \label{si:energies-intra}
    \end{subfigure}\hfill    \begin{subfigure}[t]{0.45\linewidth}
    \includegraphics[width=0.9\linewidth]{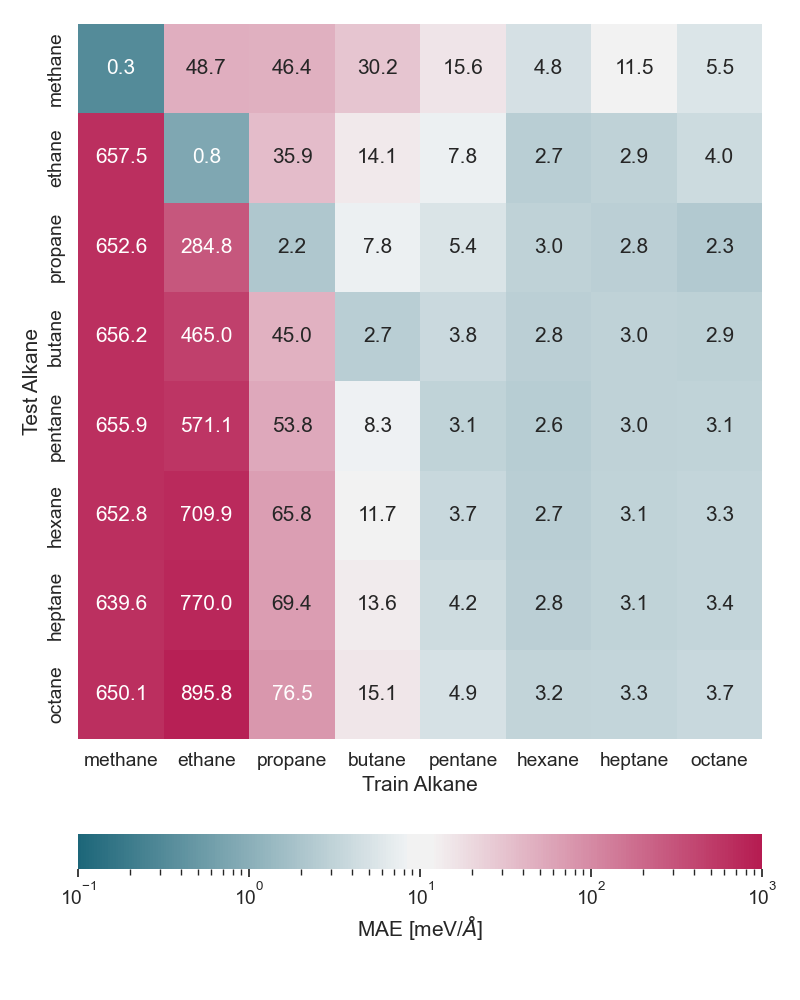}
    \caption{Force-fitting results of MACE MLIPs for different train (x-axis) and test (y-axis) pairings, respectively. Color indicates the mean-absolute-error (MAE) in units of meV/atom, respectively.}
    \label{si:forces-intra}
    \end{subfigure}
    
    \begin{subfigure}[b]{0.45\linewidth}
    \includegraphics[width=0.85\linewidth]{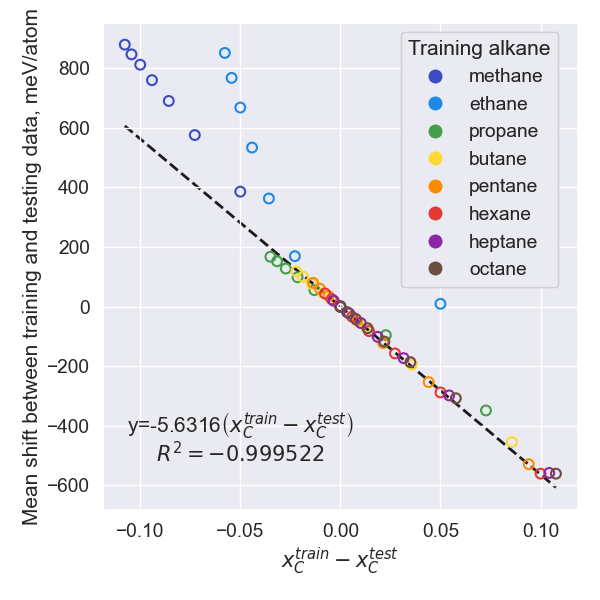}
    \caption{The relationship between the shift in composition $(x_C^{\text{train}} - x_C^{\text{test}})$ and the shift in energies for extrapolated MLIPs. The models trained on the methane, ethane, and propane datasets show a divergence from this proportionality, as they are not necessarily considered ``well-conditioned.'' Marker denotes the testing set, and color denotes the training set.}
    \label{si:mean-shift-intra}
    \end{subfigure}\hfill    \begin{subfigure}[b]{0.45\linewidth}
    \includegraphics[width=0.9\linewidth]{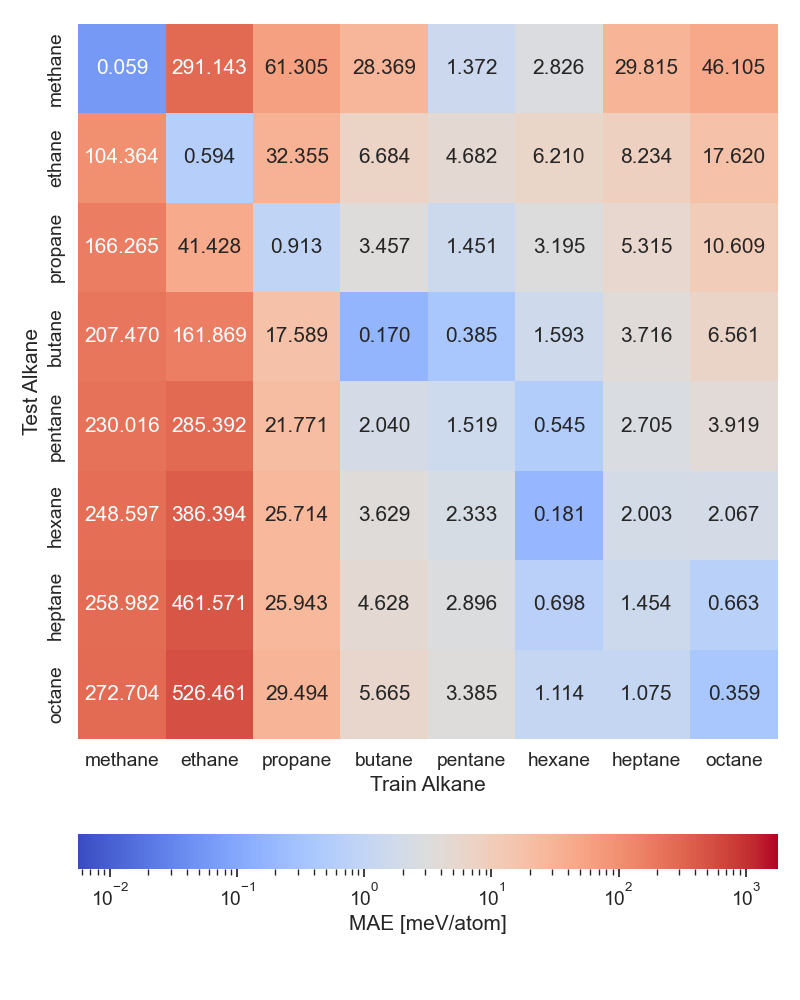}
    \caption{Shifted energy errors for MACE energies, accounting for the learnable shift in Fig.~\ref{fig:mean-shift}.}
    \label{si:mean-shift-correction-intra}
    \end{subfigure}
    \caption{Analogous to Fig.~\ref{fig:results}, results of predicting the intramolecular potential energy ($\tilde{E}_{intramolecular}=E_{intramolecular}+E_{atom}$) on $n=1-8$ alkanes.}
    \label{si:results-intra}
\end{figure*}

\begin{figure}[ht!]
    \begin{subfigure}[b]{\linewidth}
    \centering
    \includegraphics[width=0.75\linewidth]{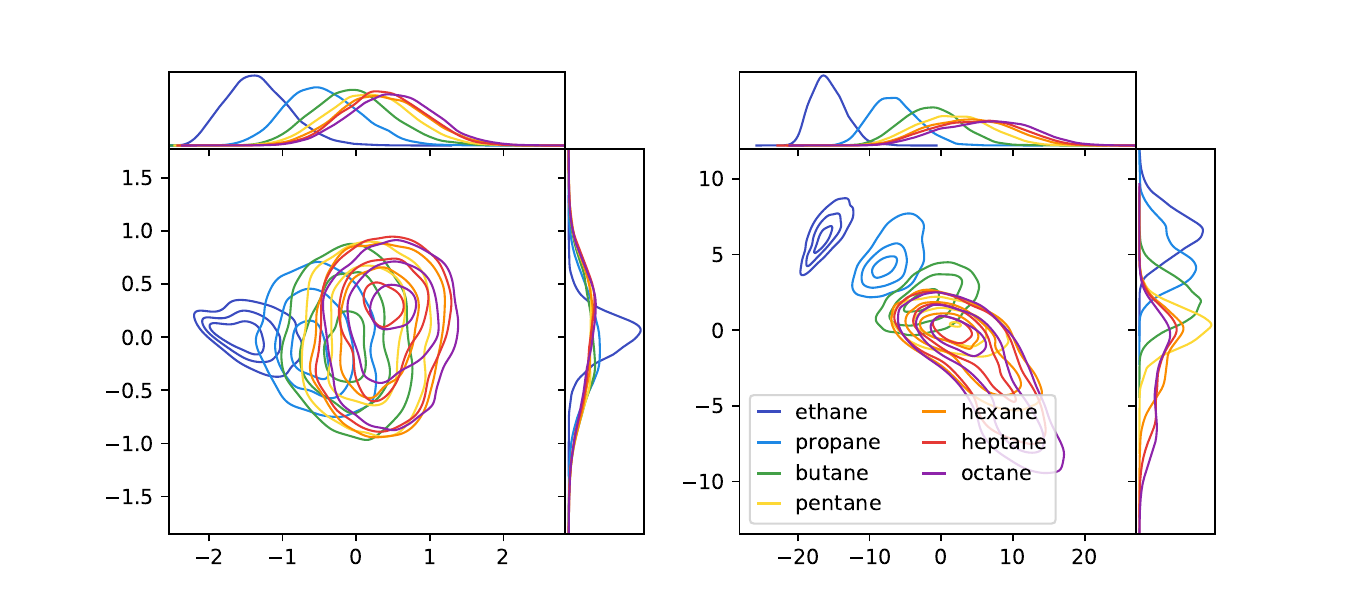}
    \caption{Maps of hydrogen environments. At a perceptive length of 10\AA, we see a convergence of hydrogen environments once reaching pentane. }
    \label{si:h}
    \end{subfigure}
    \begin{subfigure}[b]{\linewidth}
    \centering
    \includegraphics[width=0.75\linewidth]{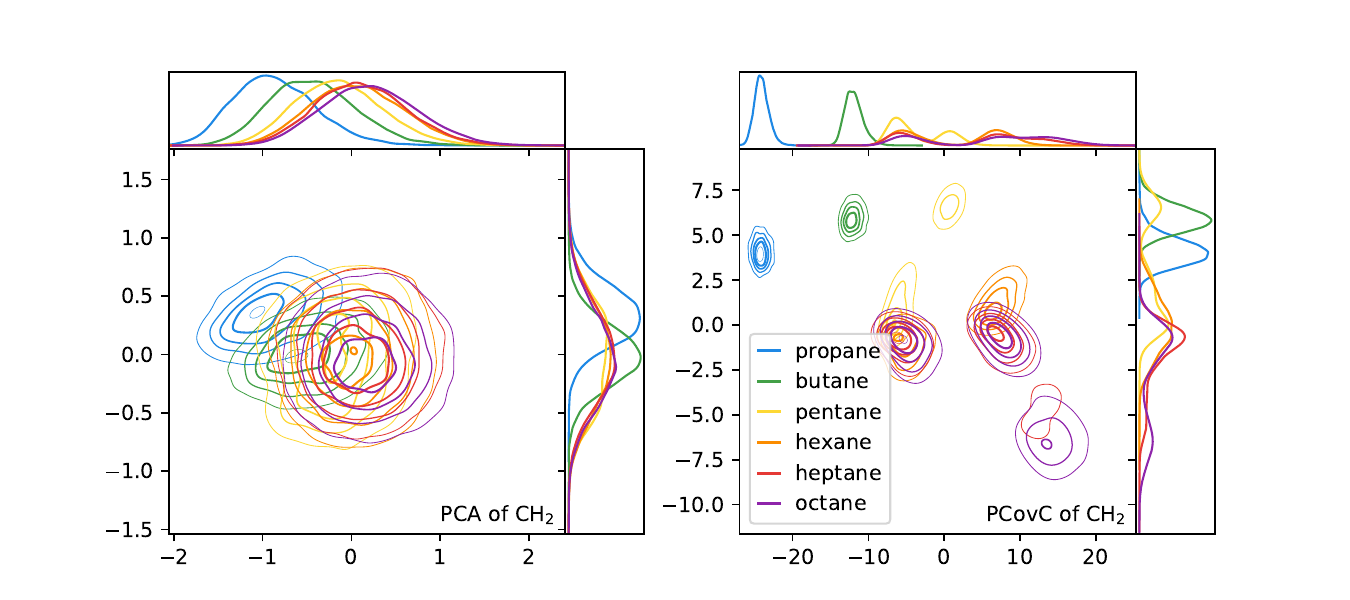}
    \caption{Maps of CH$_2$ environments. At a perceptive length of 10\AA, we see a convergence of CH$_2$ environments once reaching hexane.}
    \end{subfigure}
    \begin{subfigure}[b]{\linewidth}
    \centering
    \includegraphics[width=0.75\linewidth]{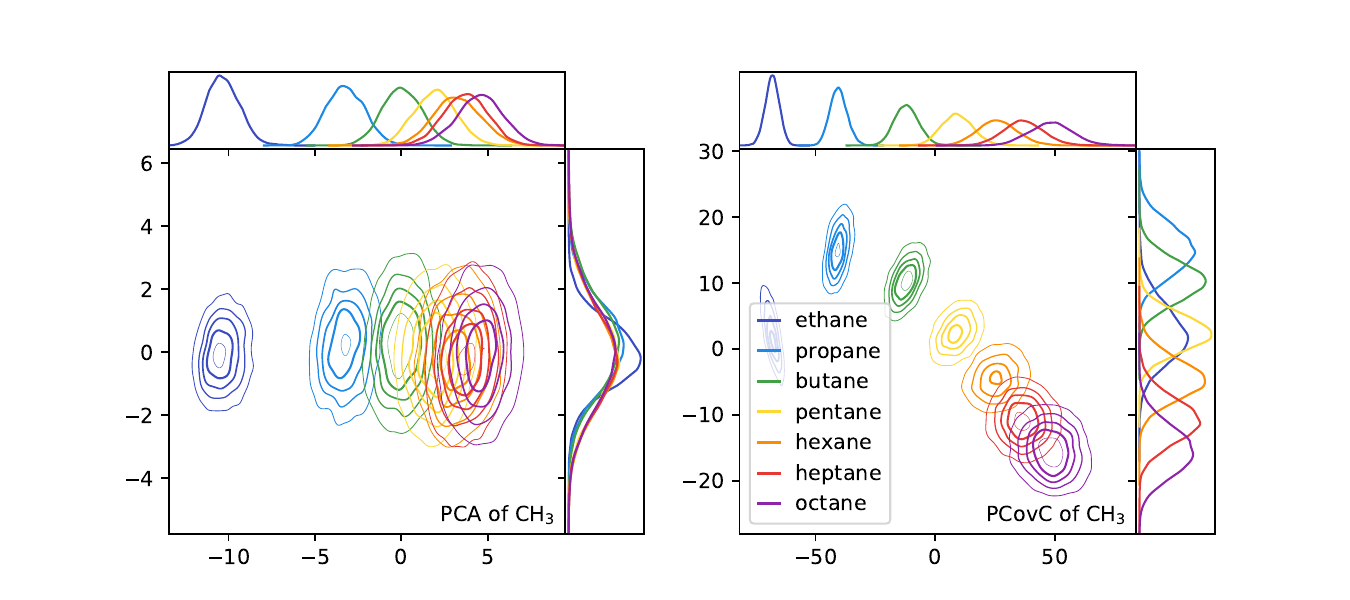}
    \caption{Maps of CH$_3$ environments. At a perceptive length of 10\AA, we see a convergence of CH$_3$ environments once reaching butane.}
    \label{si:ch2}
    \end{subfigure}
    \caption{Analogous to Fig.~\ref{fig:map}, maps of different environments for the different training sets, via Principal Components Analysis (left) and Principal Covariates Classification (PCovC\cite{jorgensen_interpretable_2025}, right)}
    \label{si:pcovc_pcova}
\end{figure}

\begin{figure}[ht!]
    \begin{subfigure}[b]{0.45\linewidth}
    \includegraphics[width = 0.9\linewidth]{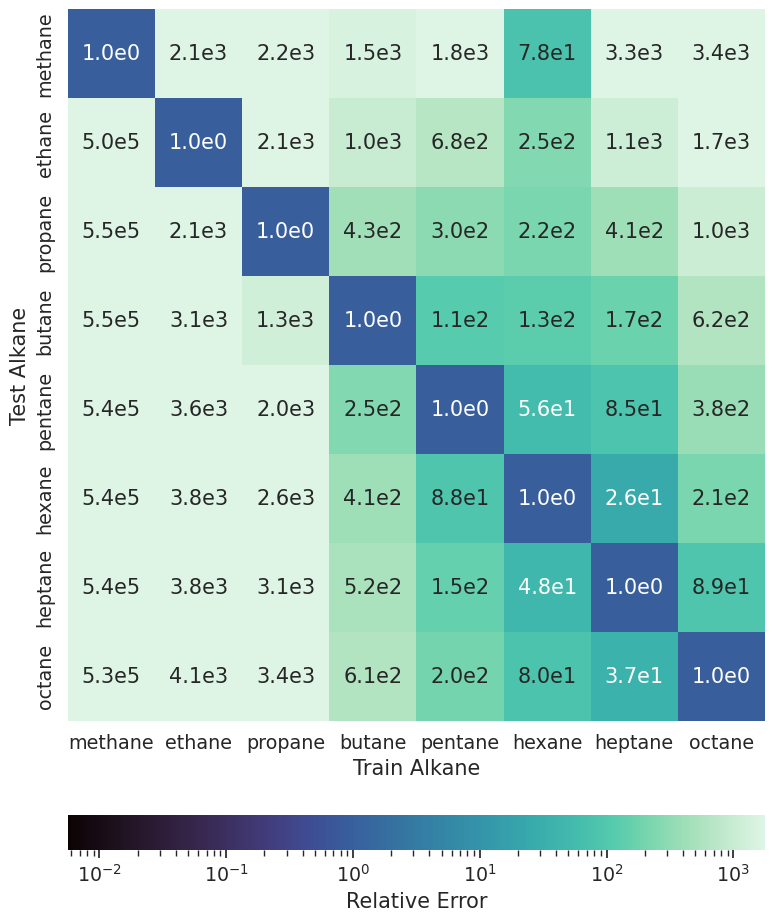}
    \caption{Relative errors of SOAP-Ridge potentials trained on $X^{total}$ and intermolecular energies.}
    \end{subfigure}\hfill    \begin{subfigure}[b]{0.45\linewidth}
    \includegraphics[width = 0.9\linewidth]{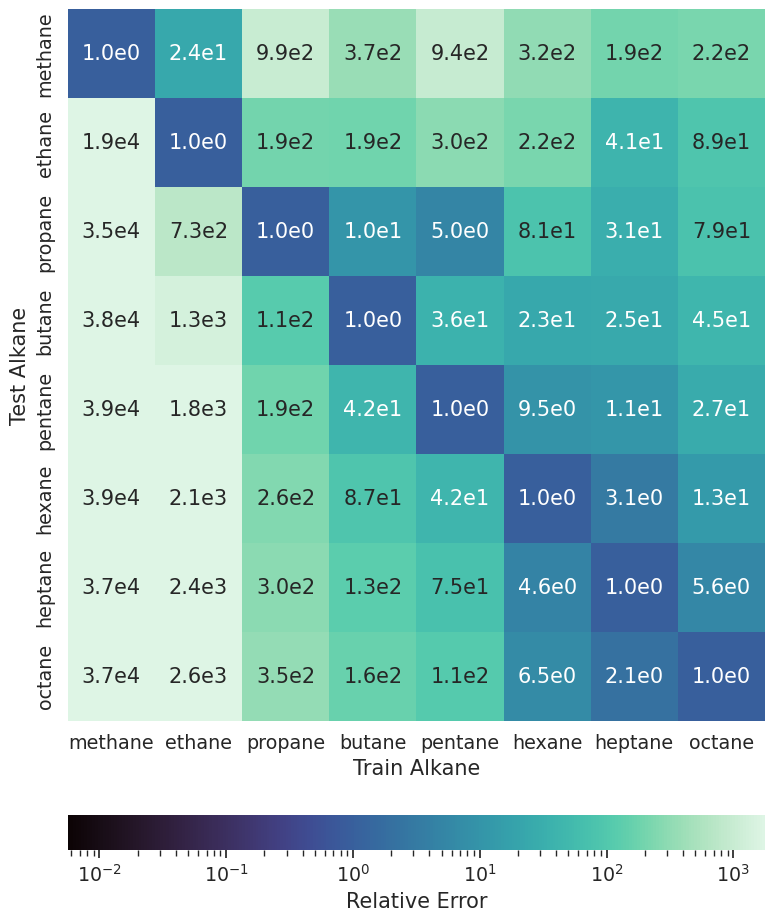}
    \caption{Relative errors of SOAP-Ridge potentials trained on $X^{fs}$ and intermolecular energies.}
    \end{subfigure}
    \caption{Analogous to Fig.~\ref{fig:soap_ridge}, heatmaps of relative errors for different training and testing sets calculated using the mean-absolute-error in meV/atom for testing sets divided by the interpolative error (error of the training alkane on its testing set).}
    \label{si:rel_soap}
\end{figure}

\begin{figure}[ht!]
    \centering
    \includegraphics[width = 0.45\linewidth]{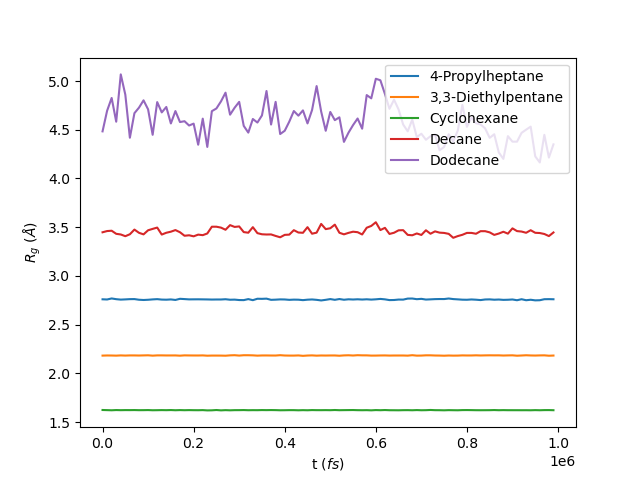}
    \caption{Mean radii of gyration for the nonlinear alkanes.}
    \label{si:rg}
\end{figure}

\end{document}